\documentclass[journal, comsoc]{IEEEtran}
\usepackage[T1]{fontenc}

\usepackage{color}
\usepackage{url}
\usepackage[square, comma, sort&compress, numbers]{natbib}
\usepackage{makecell}
\usepackage{booktabs}
\usepackage{multirow}
\usepackage{graphicx}
\usepackage{subfigure}
\usepackage{algorithmicx,algorithm}
\usepackage{enumerate}
\usepackage{algpseudocode}
\usepackage{amsmath}

\begin{document}

\title{
High Efficiency Rate Control for Versatile Video \\Coding
Based on Composite Cauchy Distribution
}

\author{Yunhao~Mao, Meng~Wang, Shiqi~Wang,~\IEEEmembership{Member, IEEE,} 
            and Sam~Kwong,~\IEEEmembership{Fellow, IEEE}\\[0.5em]        
\thanks{
Y. Mao, M. Wang, S. Wang and S. Kwong are with Department of Computer Science, City University of Hong Kong, Hong Kong, China, (e-mail: yhmao3-c@my.cityu.edu.hk; mwang98-c@my.cityu.edu.hk; shiqwang@cityu.edu.hk; cssamk@cityu.edu.hk). 
}
}
\maketitle

\begin{abstract}

In this work, we propose a novel rate control algorithm for Versatile Video Coding (VVC) standard based on its distinct rate-distortion characteristics.  
By modelling the transform coefficients with the composite Cauchy distribution, higher accuracy compared with traditional distributions has been achieved. Based on the transform coefficient modelling, the theoretically derived R-Q and D-Q models which have been shown to 
deliver higher accuracy in characterizing RD characteristics for sequences with different content are incorporated into the rate control process. Furthermore, to establish an adaptive bit allocation scheme, the dependency between different levels of frames is modelled by 
a dependency factor to describe relationship between the reference and to-be-coded frames. Given the derived R-Q and D-Q relationships, as well as the dependency factor, an adaptive bit allocation scheme is developed for  optimal bits allocation. We implement the proposed algorithm on VVC Test Model (VTM) 3.0. Experiments show that due to proper bit allocation, for low delay configuration the proposed algorithm can achieve 1.03$\%$ BD-Rate saving compared with the default rate control algorithm and 2.96$\%$ BD-Rate saving compared with fixed QP scheme. Moreover, 1.29$\%$ BD-Rate saving and higher control accuracy have also been observed under the random access configuration.
\end{abstract}


\IEEEpeerreviewmaketitle

\section{Introduction}

\IEEEPARstart{W}{ith} the widespread of multimedia services, recent years have witnessed an explosive increase of video data, bringing grand challenges to video data management in terms of storage and transmission. The video coding standards which have evolved for several decades from H.264/AVC~\cite{wiegand2003overview}, H.265/HEVC~\cite{sullivan2012overview} to the emerging Versatile Video Coding (VVC)~\cite{L1001-v9} standard, have been repeatedly proven to improve the coding efficiency beyond the previous one. 
A series of novel video coding technologies have been investigated during the standardization of VVC, aiming at providing more efficient video compression solutions. 
To better adapt the characteristics of high resolution videos, the size of coding tree unit (CTU) is enlarged to 128 $\times$ 128 with the cooperation of more flexible partitions such as quad-tree, binary-tree and ternary-tree~\cite{MTT}. Besides, enhanced intra and inter prediction technologies~\cite{HMVP_DCC,ISP_VVC,AMC,AMVR,WideAngular} are investigated to further remove the spatial and temporal redundancies. 
Moreover, multiple transform selection (MTS) is supported for better compacting residual energies~\cite{Transform_Zhaoxin} in frequency domain. Regarding quantization, dependent quantization is adopted, which maps quantization candidates within one block into a trellis map. The path with the lowest rate-distortion (RD) cost is determined as final quantization outcomes~\cite{Trellis_DCC}.

As an essential component of an encoder, rate control, which has been widely investigated since MPEG-2~\cite{MPEG2}, aims to provide the best video quality with the constraint of bit-rate budget.
Rate control is crucial for real-application scenarios of the video codec with the regularization of the bit-rate. Generally speaking, there are two main procedures in rate control: bit-rate allocation and coding parameter determination. Bit-rate allocation can be processed with three-levels: the group of pictures (GOP) level, frame level, and CTU level.
With GOP level bit allocation, the encoder assigns available bits to the to-be-encoded GOPs with the consideration of buffer occupancy. In a GOP, bits are allocated to each frame based on GOP structure~\cite{li2014lambda} or pre-analyzed RD characteristics~\cite{li2016lambda}. In the literature, there are two ways to realize frame-level bit allocation: fixed ratio allocation~\cite{li2014lambda} and adaptive ratio allocation~\cite{li2016lambda}. More specifically, fixed ratio bit allocation generally utilizes a predefined ratio depending on frame structure and target bit-rate. In~\cite{li2016lambda}, the authors proposed an adaptive bit allocation algorithm for HEVC based on $\lambda$ domain rate control. The adaptive bit allocation algorithms are mostly built on an RD model, 
and the bit-rate control is realized by modelling the relationship among the rate, distortion, and coding parameters, where the coding parameters could be the Lagrange multiplier $\lambda$, the quantization parameter $QP$ {(or quantization step size $Q$)} and the percentage of zero coefficients $\rho$~\cite{zhihai2001low}.

Existing rate control algorithms attempt to exploit the relationship among $QP$, target bit-rate $R$ and $\lambda$. However, most of them merely focus on establishing an elaborately designed relationship between $R$ and $QP$ or $R$ and $\lambda$. 
In particular, $Q$-domain rate control algorithms emphasize on the importance of $QP$ whereas ignoring the role of $\lambda$, which is decisive in mode decision. Moreover,
in the sense of $\lambda$-domain rate control, $QP$ is no longer the most critical factor.
$\lambda$-domain rate control shows the advantage over $Q$-domain rate control in HEVC encoder, which collaborates well with more sophisticated mode selection schemes.
Although $\lambda$ plays an important role in mode decision, the influence of $\lambda$ on output distortion and bit-rate is still quite obscure. By contrast, $QP$ influences both the mode decision and quantization outcomes which dominate coding distortions and bit-rate. 
This inspires us to construct a new analytical framework incorporating with $R$, $Q$ and $\lambda$ to better capture the inner-connections among these three. For computational convenience, we employ quantization step size $Q$ in the proposed model, which can be monotonously mapped from $QP$.

The rate control philosophy in VVC inherits from H.265/HEVC with minor modifications for attending the {ever increasing SKIP coded blocks~\cite{K0390}}. As more advanced technologies are adopted in the VVC, the RD characteristics as well as the $QP$ and $\lambda$ relationship become more flexible. To further promote the rate control efficiency for VVC, in this paper, we first propose to model the distribution of transform coefficients with an improved discrete Cauchy distribution that could more accurately depict the behavior of transform coefficients. 
Subsequently, we explore a new relationship among $Q$, coding bits and distortions based on the discrete Cauchy distribution model. Moreover, an optimal bit allocation scheme at GOP-level and frame-level is proposed in an analytical way by leveraging the reference dependencies in terms of distortions and coding bits. In this manner, better RD performance can be achieved with the proposed rate control scheme. Extensive experimental results show that the proposed scheme can achieve 1.03\% and 1.29\% BD-Rate savings  compared  with  the  default  rate  control algorithm in VTM platform~\cite{K0390} in low-delay B (LDB) and random-access (RA) configurations.

\section{Related Works}
Existing rate control algorithms~\cite{li2014lambda,zhihai2001low,ma2005rate,8694021} strive to achieve more precisely modelling of the relationships between coding parameters and bit-rate, with the aim of capturing the RD characteristics in different video sequences. The most intuitive way to obtain a robust relationship is to encode the sequence for multiple rounds with different $QP$s. However, this significantly elevates encoding complexity, making it impracticable in one-pass or two-pass coding scenarios. 
Coding distortion $D$ is mainly introduced by quantization, and the number of output bits $R$ is closely related to the entropy coding of quantized residuals. As such, it is feasible to model the RD behavior according to the distribution of transform coefficients.
\subsection{Distribution of Transform Coefficients}
In the literature, numerous models have been investigated to model the distribution of transform coefficients.
In \cite{Gish1968}, source codes are modelled with uniform distribution within each quantization interval. Cooperating with hard quantization process, a quadratic relationship between quantization step size $Q$ and distortion $D$ can be obtained as follows,
\begin{equation}
\label{1.1}
    D=\frac{Q^2}{12}.
\end{equation}
However, it is widely acknowledged that coefficient distribution may not be subject to the uniform distribution in real application scenarios, and such assumption only holds under high bit-rate conditions~\cite{seo2013rate}.
Besides, a series of classical distribution models such as Gaussian distribution, Laplacian distribution and Cauchy distribution have been studied in the literature \cite{muller1993distribution, eude1994distribution, lam2000mathematical, yang2014transparent}. Gaussian distribution reveals the advantage in parameter estimation but with poor accuracy in fitting actual distribution~\cite{li2009laplace,cui2017hybrid,seo2013rate}. 
Generalized Gaussian distribution can properly model the coefficients distribution whereas the associated controlling parameters are difficult to estimate. Laplacian distribution has been widely employed in video coding tasks, as it strikes an excellent trade-off between the fitting accuracy and computational complexity regarding the parameter estimation.
In \cite{li2009laplace}, Li \textit{et al.} modelled residuals with Laplacian distribution and derived close-forms for $R$-$Q$ and $D$-$Q$ expression, by which a better $\lambda$ is inferred for rate-distortion optimization (RDO), bringing 1.60 dB gains on average in terms of PSNR.  In \cite{cui2017hybrid}, a low-complexity rate distortion optimized quantization (RDOQ) scheme is investigated based on a hybrid Laplacian distribution modelling for HEVC. Moreover, Seo \textit{et al.}~\cite{seo2013rate} proposed a rate control algorithm based on Laplacian distribution aiming at minimizing video quality fluctuation.  In~\cite{kamaci2005frame}, it was observed that Cauchy distribution can more accurately model the distribution of the AC coefficients than Laplacian distribution whereby a frame level bit allocation scheme is investigated for H.264/AVC.
\subsection{Rate Control}
In rate control, efforts have been devoted to establishing the relationship among $QP$,  $R$ and $\lambda$. These methods operate in \(\rho\) domain, \(Q\) domain and \(\lambda\) domain to regularize the coding bit-rate.  

Typically, \(\rho\) domain methods ~\cite{zhihai2001low} assume a linear relationship between coding bit-rate $R$ and the percentage of zero coefficients $\rho$,
\begin{equation}
    R=\theta(1-\rho),
\end{equation}
where $\theta$ is a parameter relevant to the video content. 
As such, a one-to-one mapping between $R$ and $QP$ can be derived with the assistant of the intermediate $\rho$. Even though $\rho$-domain rate control could provide smoother output bit-rates and better objective quality, it was designed for H.263 targeting at coping with fixed block size, which may impede its further applications. 

In~\cite{ma2005rate}, a complexity-adjustable rate control scheme based on a reliable $R$-$Q$ relationship was investigated for H.264/AVC. More specifically, a linear relationship between $R$ and $Q^{-1}$ is observed,
\begin{equation}
    R =\frac{Z\cdot SAD}{Q} + r_h.
\end{equation}
where $SAD$ denotes the sum of absolute difference of the motion-compensated micro-block. $Z$ and $r_h$ represent model parameter and the number of header bits, respectively. Typically, they are highly related to the slice type. Comparing with the fixed QP configuration, this rate control algorithm achieves 0.33 dB PSNR gain with negligible coding time increase.

Regarding the $\lambda$ domain rate control, the hyperbolic function based RD relationship, which is recognized to hold better fitting accuracy~\cite{ardestani2010rate} than the conventional exponential function~\cite{sullivan1998rate}, is employed in HEVC~\cite{li2014lambda}. The relationship between $R$ and $D$ can be formulated as follows,
\begin{equation}
\label{1.2}
    D(R)=UR^{-V},
\end{equation}
where $U$ and $V$ are model parameters.
Moreover, the RD cost $J$~\cite{everett1963generalized} can be described as,
\begin{equation}
\label{1.3}
    J=D+\lambda R.
\end{equation}
When encoding a sequence, a set of coding parameters which can minimize $J$ is preferable. To find the best bit-rate which can minimize $J$, the derivative of $J$ with respect to \(R\) is calculated and set to zero as follows,
\begin{equation}
\label{1.4}
    \frac{\partial J}{\partial R}=\frac{\partial D}{\partial R}+\lambda = 0.
\end{equation}
With the combination of Eqn.~(\ref{1.2}), the relationship between $\lambda$ and $R$ can be obtained as follows,
\begin{equation}
\label{1.5}
    \lambda=-\frac{\partial D}{\partial R}=\mu R^\varphi,
\end{equation}
where $\mu$ and $\varphi$ are model parameters which are closely relevant to video content. In~\cite{li2014lambda}, a parameter updating strategy is employed, with which $\mu$ and $\varphi$ can be updated synchronously in the coding process. In this manner, given the target bit rate, the corresponding $\lambda$ can be obtained through the $\lambda$-$R$ relationship in Eqn.~(\ref{1.5}). 
Moreover, the associated QP can be derived according to a linear transform with $\ln\lambda$~\cite{li2013qp},
\begin{equation}
\label{1.6}
    QP=4.2005\cdot\ln\lambda+13.7122.
\end{equation}

To further improve the performance, a $\lambda$-domain adaptive bit allocation scheme is investigated~\cite{li2016lambda} for HEVC rate control. By exploring the inter frame dependency, two hypothesises are raised, including the linear relationship regarding the distortions between reference and current frames, and low dependencies regarding the frame-level bits between reference and current frames. Subsequently, an optimal bit allocation scheme cooperated with a predefined ratio is proved to be more effective than fixed allocation ratio. 

In~\cite{K0390}, a new parameter estimating strategy for $\lambda$ domain rate control is proposed and adopted by VVC.
The $\lambda$ used by the previous encoded frame at the same temporal layer is regarded as the optimal one for the current frame. As such, the RD relationship can be predicted according to specific RD point and corresponding slope $\lambda$. 
Though traditional $\lambda$ domain rate control schemes adopted as a reference in VVC show promising RD performance and stable output bit-rate, the $R$-$\lambda$ and $\lambda$-$Q$ relationship built upon parameter estimation may not be able to fully adapt the properties of video content without the thorough consideration of transform coefficients. Considering the fact that RD performance is highly related to transform coefficients, we propose a distribution based rate control algorithm. The distribution of transform coefficients is modelled with an improved discrete Cauchy distribution. Based on the proposed model, the R-Q and D-Q models that are built upon the characteristics of the video content are derived for encoding parameter estimation. 

\section{Cauchy Distribution Based Transform Coefficient Modelling}
\begin{figure}[t]
\centering
\subfigure[]{   
\includegraphics[scale=0.5]{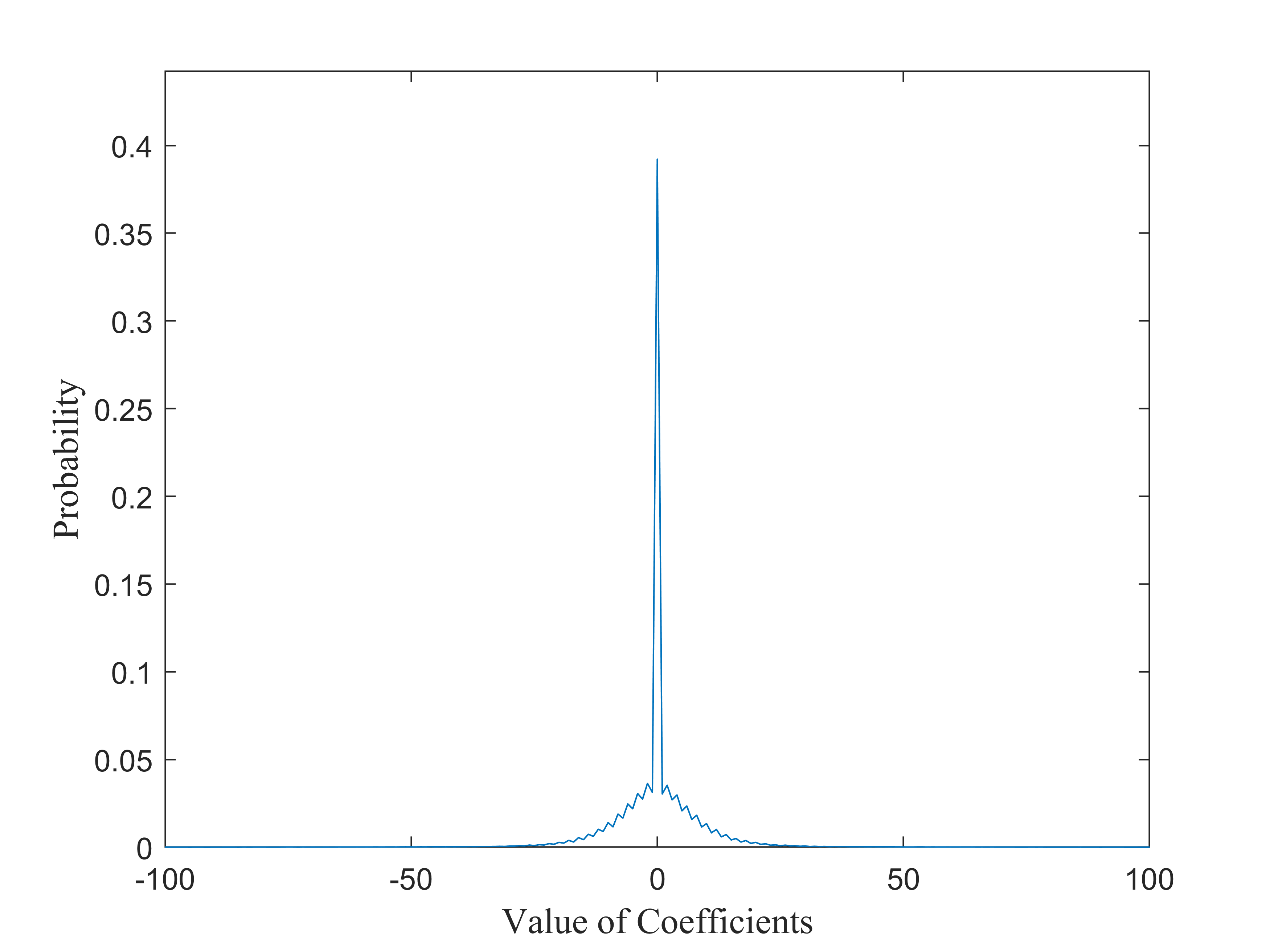}}
\subfigure[]{   
\includegraphics[scale=0.5]{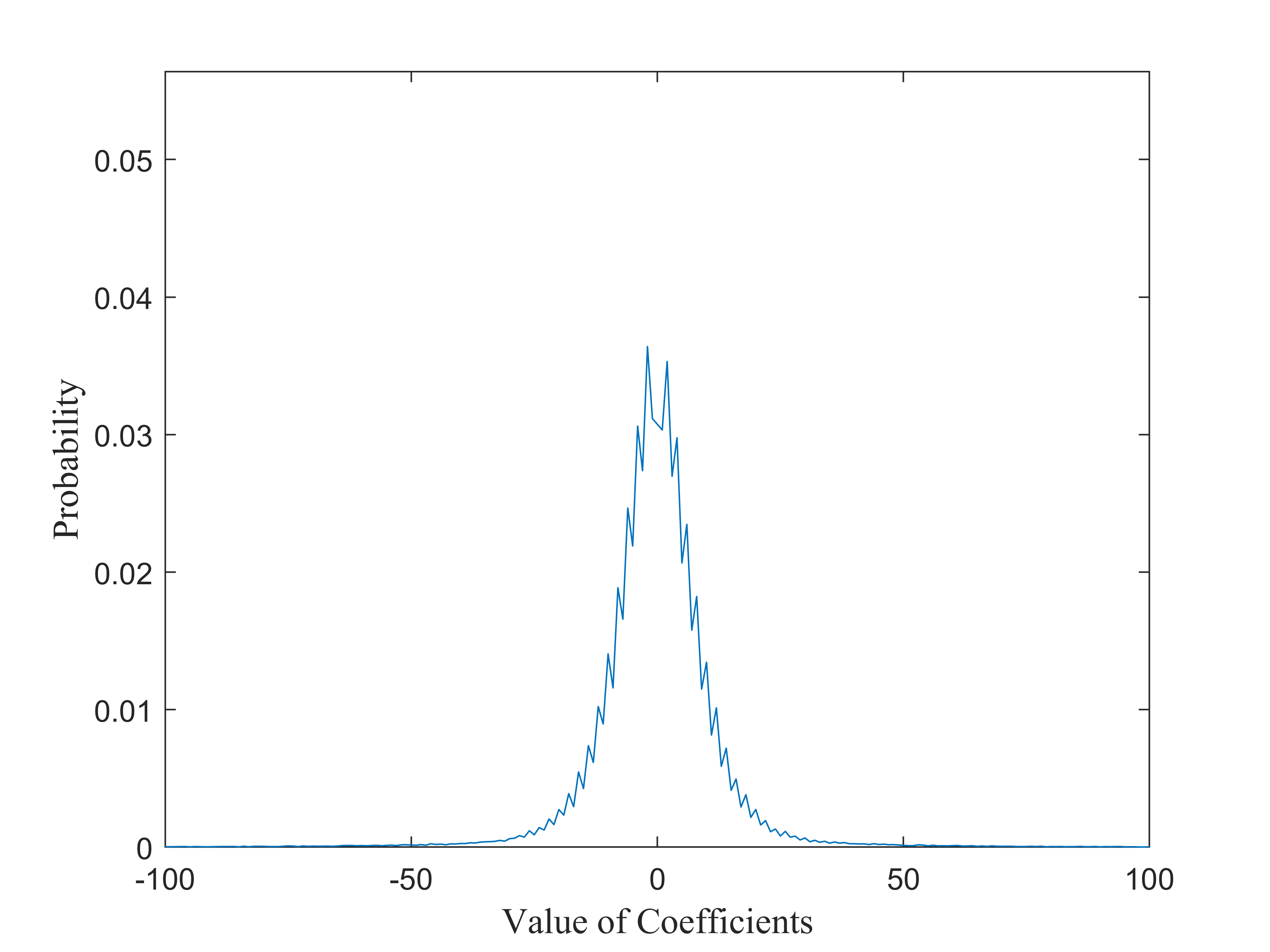}}
\caption{Distribution of transform coefficients of a B frame in sequence ``BasketballDrill'' under LDB configuration. (a) Zero included; (b) Zero excluded.}
\label{fig:distribution}
\end{figure}

In this section, we establish a new model that exhibits high accuracy in characterizing the transform coefficients in VVC, serving foundation to describe the 
the relationship between R-D and coding parameters. 
It is widely acknowledged that the transform coefficients exhibit a symmetrical distribution with peak at zero. 
Fig.~\ref{fig:distribution} shows the distribution of the transform coefficients of a typical B frame 
from sequence ``BasketballDrill'', wherein the inclusion and exclusion of zero point are respectively illustrated. 
We can observe a symmetric distribution with a peak locating at the zero point, and the distribution decreases rapidly as the coefficients deviate from zero. 
Such peaking at zero motivates us to develop a composite distribution that 
models the zero and non-zero coefficients separately, in an effort to achieve higher fitting accuracy.

Previous research~\cite{altunbasak2004analysis} indicates that Cauchy distribution is efficient in approximating the distribution of DCT coefficients. 
In the proposed distribution, we adopt a composite modelling strategy based upon the peaking zero and discrete Cauchy distribution for non-zero coefficients, 
\begin{equation}
\label{2.1}
    \rho(n)=\begin{cases}
    \frac{\alpha}{n^2+\beta},\quad &n\in Z \cap n \neq 0,\\
    p_0,\quad &n=0,
    \end{cases} 
\end{equation}
where $\alpha$ and $\beta$ are distribution parameters. $p_0$ is the probability of zero coefficient, and $n$ denotes the coefficient level. Considering that involving zero coefficients in the distribution may cause a local minimum during parameter estimation, the proposed distribution typically excludes the inferences of zero point to ensure higher accuracy for non-zero parts.  

Since the sum of the proposed probability model equals to one, the inherent relationship between $\alpha$ and $\beta$ can be derived as follows,
\begin{equation}
\label{2.4}
   \begin{aligned}
   &\sum_n \rho(n)=1-p_0, n\in Z \cap n \neq 0, \\
    &\sum_{n=1}^{\infty} \frac{\alpha}{n^2+\beta} =\frac{1-p_0}{2},\\
    &\alpha =\frac{(1 - p_0)\cdot\beta\cdot\tanh(\beta^{0.5}\cdot\pi)}{(\beta^{0.5}\cdot\pi - \tanh(\beta^{0.5}\cdot\pi))}.
    \end{aligned}
\end{equation}
In practical implementation, the parameter $\beta$ is obtained by searching within a given range, targeting at minimizing the mean squared error between the modelled and actual distribution of transform coefficients.

\begin{figure*}[t]
\centering
\subfigure[]{   
\includegraphics[width=1.7in]{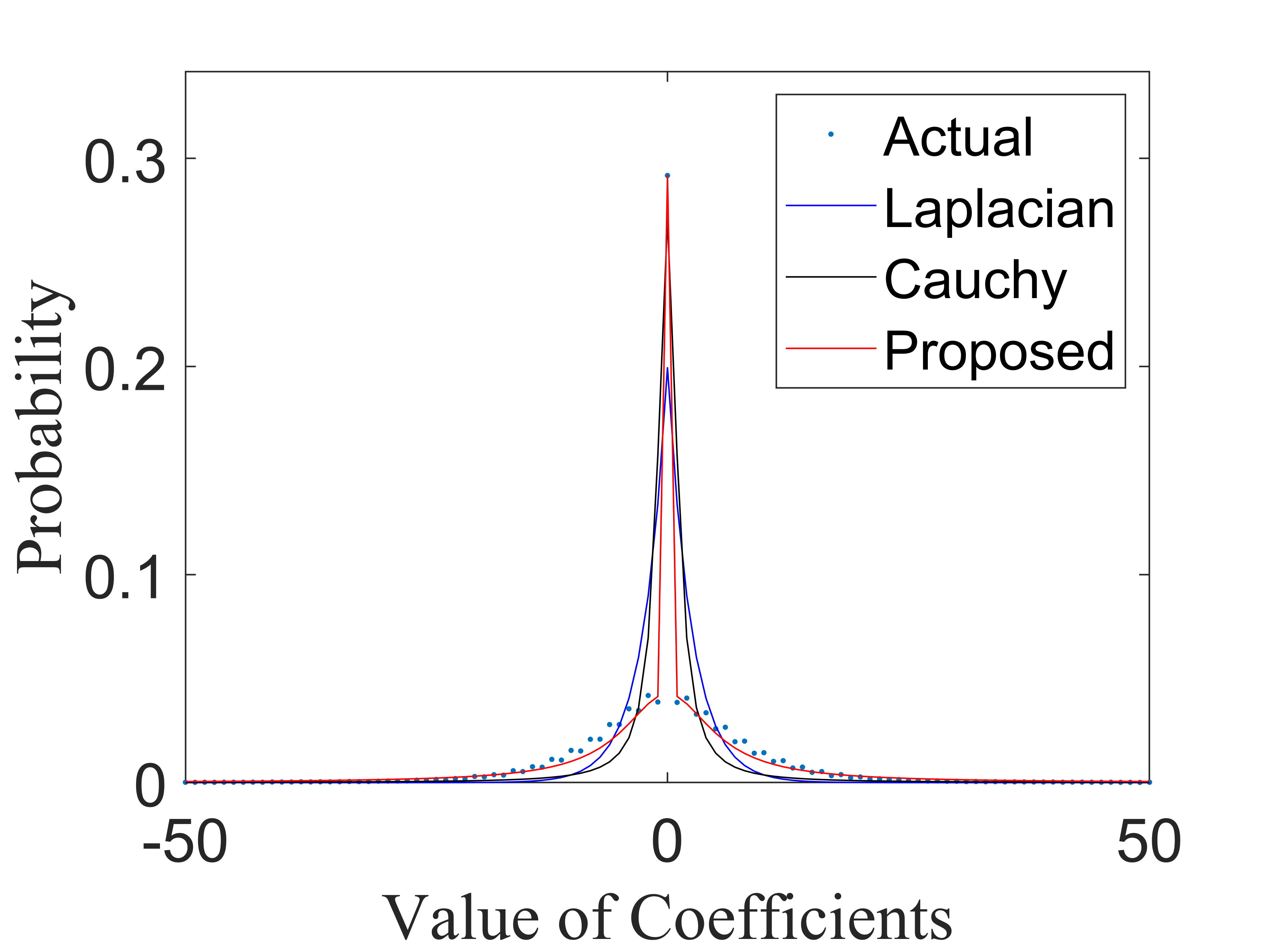}}
\subfigure[]{   
\includegraphics[width=1.7in]{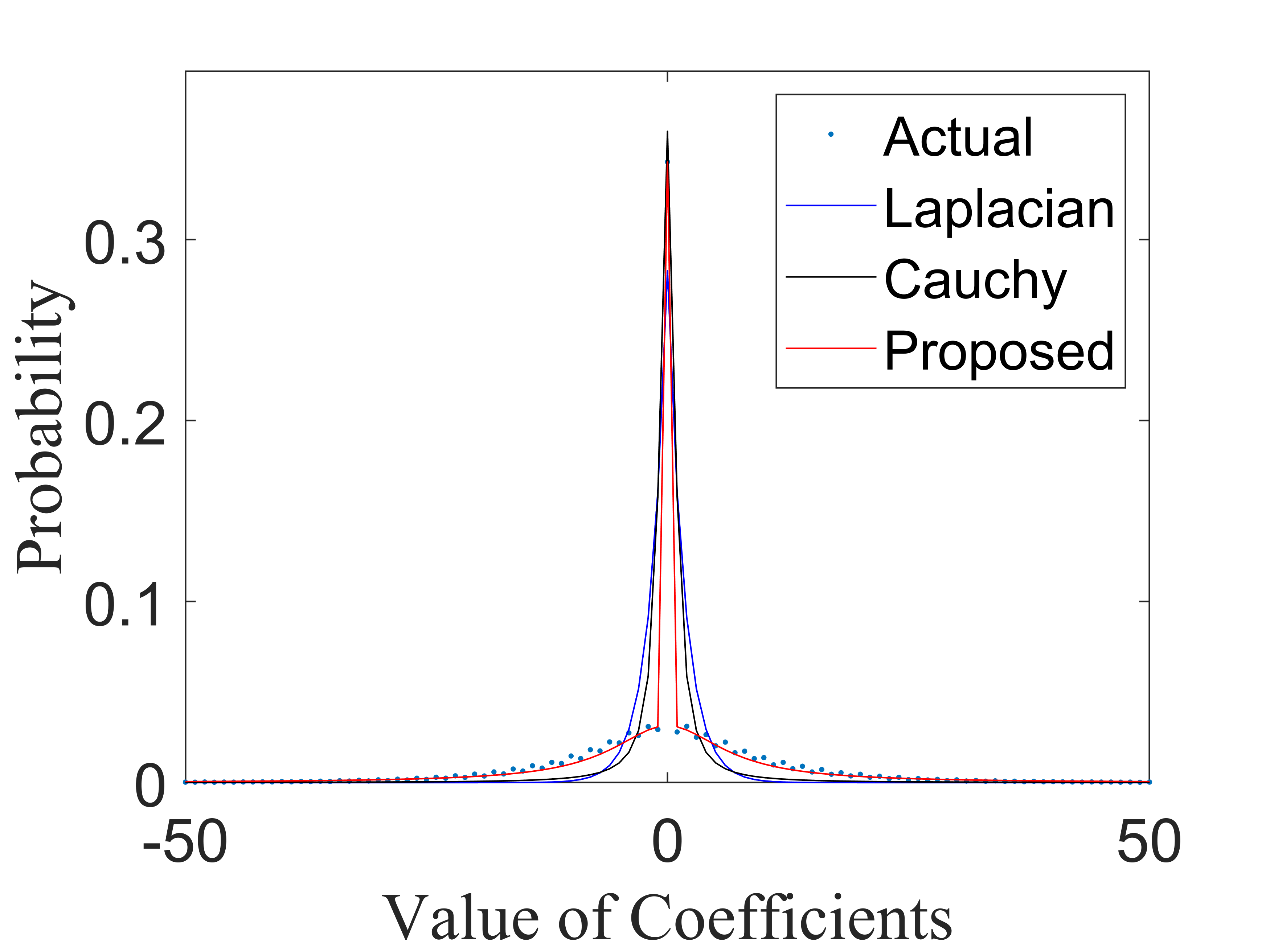}}
\subfigure[]{   
\includegraphics[width=1.7in]{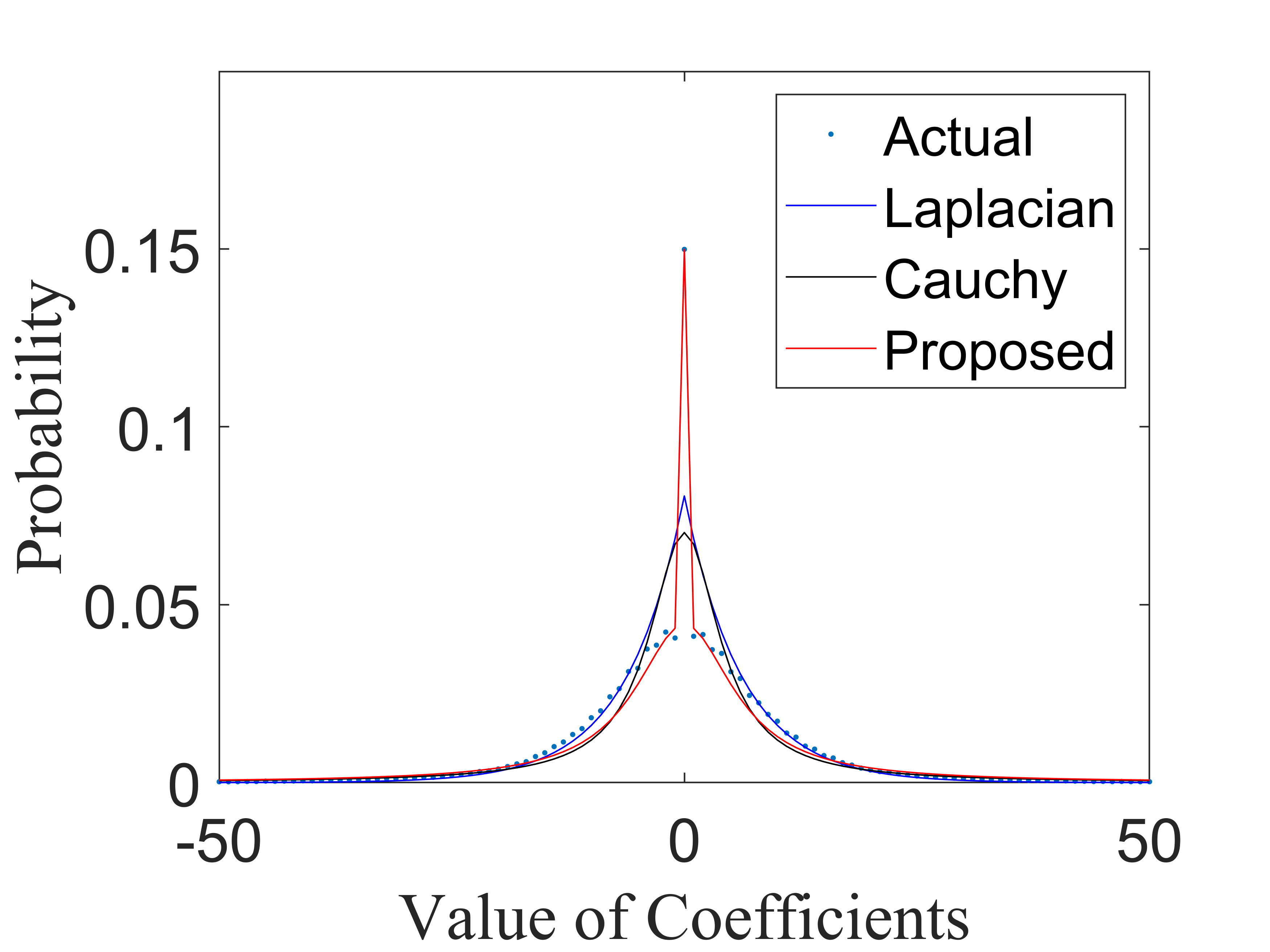}}
\subfigure[]{   
\includegraphics[width=1.7in]{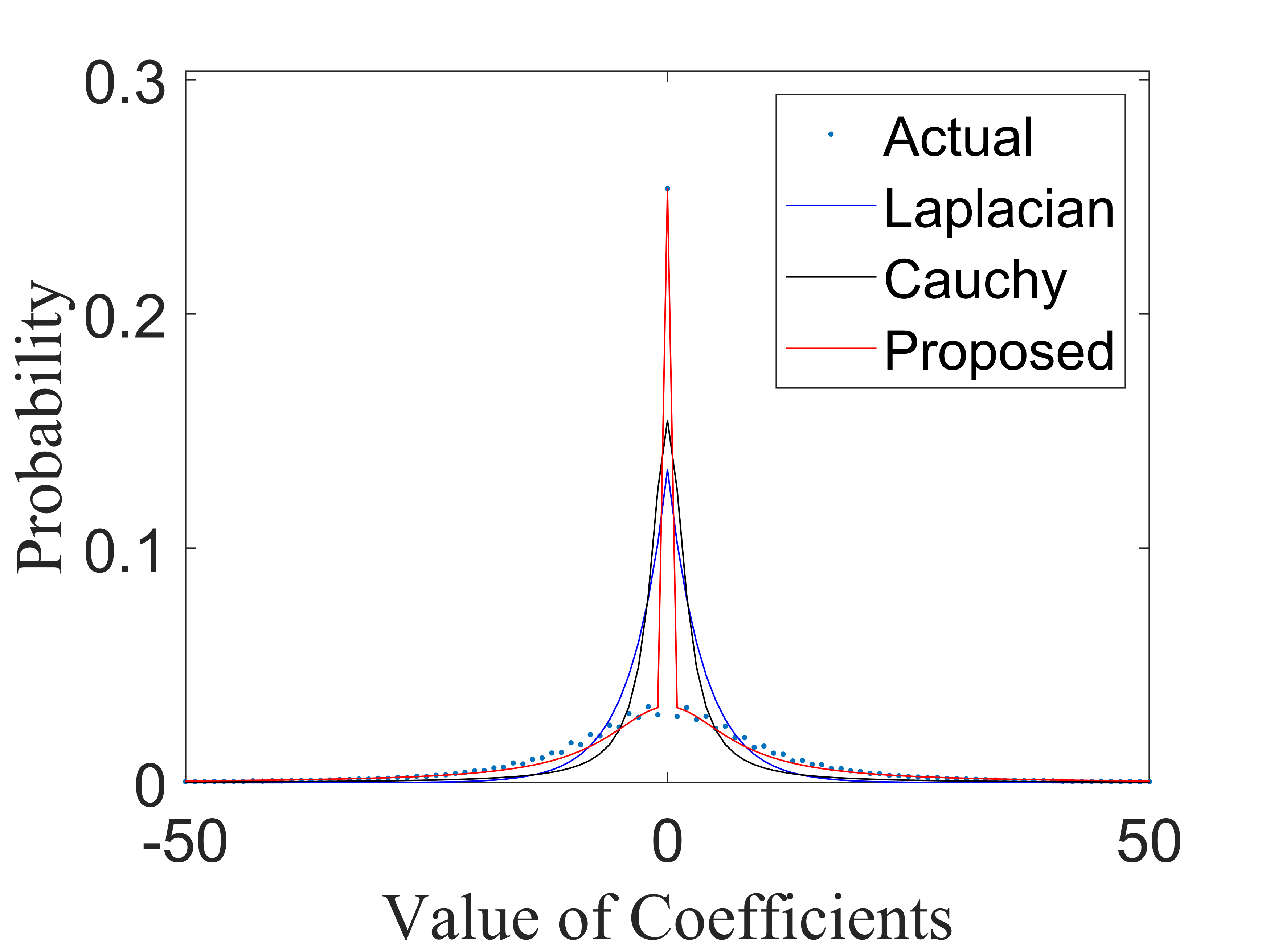}}
\caption{Comparisons among actual, Laplacian, Cauchy and the proposed composite distribution for sequences ``BasketballDrill'' and ``BQMall''. (a) ''BasketballDrill'', $QP=23$. (b) ''BasketballDrill'', $QP=28$. (c) ''BQMall'', $QP=23$. (d) ''BQMall'', $QP=28$.}\label{fig:fittingdistribution}
\end{figure*}

We compare the proposed model with Laplacian distribution and traditional Cauchy distribution regarding the fitting accuracy where Kullback-Leibler (KL) divergency~\cite{kullback1951information} is used. Given an actual coefficient distribution $f_{r}$ and statistical model $f_{p}$, the associated KL divergency can be calculated as follows,
\begin{equation}
\label{2.5}
    \mathcal{D}_{KL}=\sum_{i}f_r(n)\log\left(\frac{f_r(n)}{f_p(n)}\right).
\end{equation}
Video sequences ``BasketballDrill'' and ``BQMall'' are involved in the analyses with LDB configuration. Transform coefficients in the 16-th frames are extracted from those two sequences.
The corresponding KL divergencies are shown in Table \ref{tbl:distribution}. It can be observed that compared with the traditional distributions, the proposed model achieves higher fitting accuracy for non-zero parts, as the KL divergency between raw data and the proposed model is much lower than that of traditional models.
Fig.~\ref{fig:fittingdistribution} illustrates the comparisons among the three distribution models, and it can be noticed that the proposed model could better handle the zero-level and non-zero coefficients.

\begin{table}[t]
  \centering
\caption{Comparison of $\mathcal{D}_{KL}$ regarding Laplacian, Cauchy, and the Proposed Discrete Cauchy distribution.}
\label{tbl:distribution}
\begin{tabular}{cccc}
\toprule
Sequence& Laplacian&Cauchy&Proposed\\
\midrule
BasketballDrill, QP=23& 0.7923&0.3224&0.0591\\
BasketballDrill, QP=28& 2.0977&0.4552&0.0465\\
BQMall, QP=23& 0.1286&0.1067&0.0677\\
BQMall, QP=28& 0.8162&0.3144&0.0461\\
\bottomrule 
\end{tabular}
\end{table}

\section{Rate and Distortion Models}
In this section, we develop an analytical framework to explore the relationships among rate, distortion and coding parameters based upon the proposed composite coefficient distribution model. In particular, the $R$-$Q$ and $D$-$Q$ models are developed, serving as the foundation of the proposed rate control scheme. 
\subsection{$R$-$Q$ Model}
\begin{figure}[t]
\centering
\includegraphics[scale=0.5]{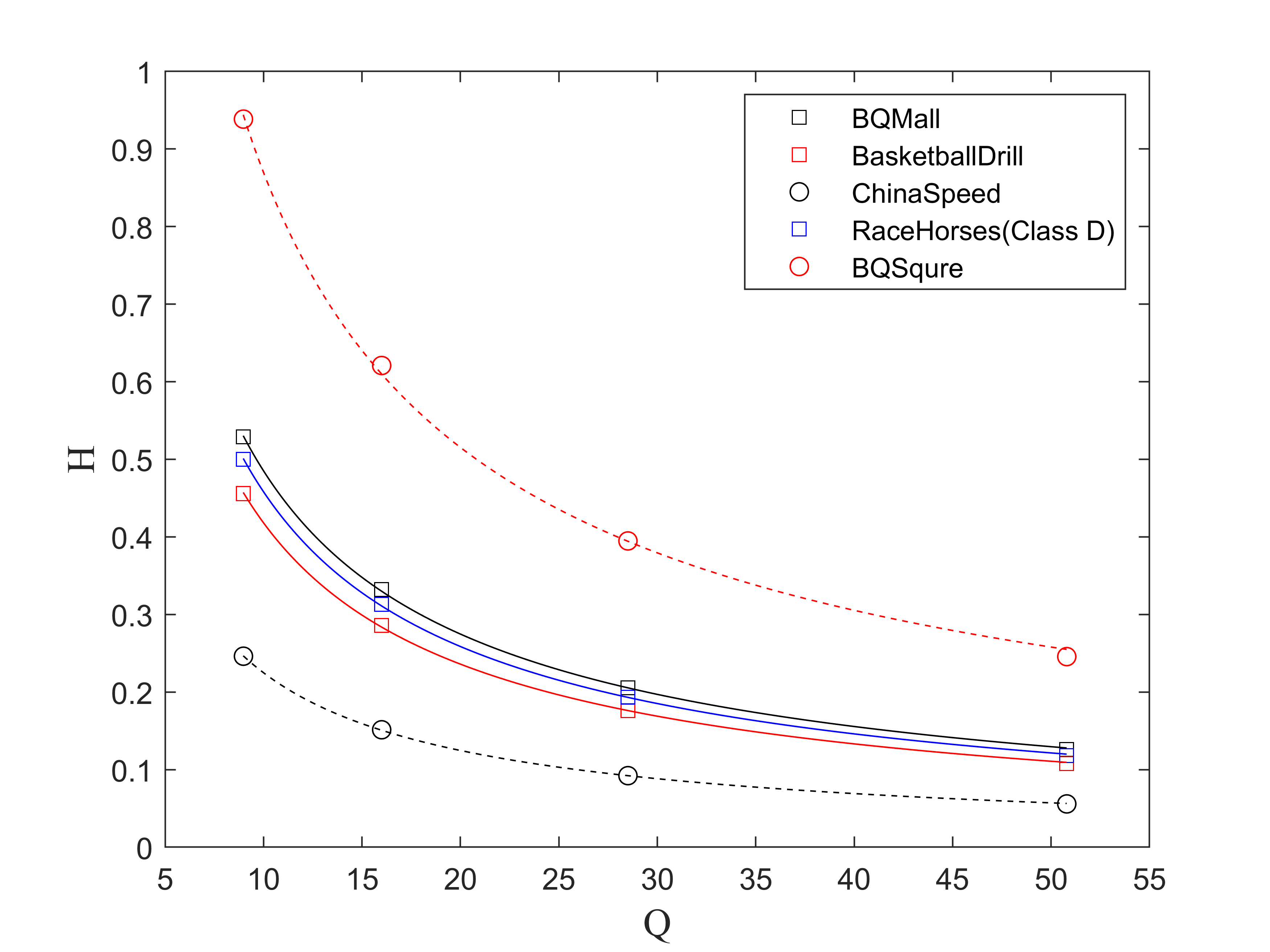}
\caption{Illustration of the relationship between quantization step size and estimated entropy of residuals.}\label{fig:relationshipQH}
\end{figure}

Herein, we utilize hard-decision quantization to simulate the dependent quantization process for simplicity~\cite{li2009laplace}. 
Given the transform coefficient $c$ and quantization step size $Q$, the quantization level $l$ can be derived as,
\begin{equation}
\label{3.2}
    l=floor\left(\frac{c}{Q}+\gamma\right),
\end{equation}
where $\gamma$ is the rounding offset which equals to $\frac{1}{3}$ for I-slice and $\frac{1}{6}$ for B-slice and P-slice~\cite{HDQ}. According to the coefficient distribution model in Eqn.~(\ref{2.1}), the probability of the $N$-th quantization level can be calculated as follows,
\begin{equation}
\begin{split}
\label{3.3}
    &P_N(Q)=\sum_{n=\lfloor(N\cdot Q-\gamma Q)\rfloor}^{\lfloor(N+1)Q-\gamma Q\rfloor}\frac{\alpha}{n^2+\beta}, \\
    &P_{-N}(Q)=\sum_{n=\lfloor(-(N+1)Q+\gamma Q)\rfloor}^{\lfloor-N\cdot Q+\gamma Q\rfloor}\frac{\alpha}{n^2+\beta},\\
    &P_0(Q)=1-2\cdot\sum_{N=1}^{L_{max}} P_N(Q),
\end{split}
\end{equation}
where $L_{max}$ is the maximum quantization level and $N$ is an integer number which ranges from 1 to $L_{max}$.
For the convenience of calculation, definite integral can be used to approximate $P_N$ as follows,
\begin{equation}
\begin{split}
\label{3.5}
   &P_N(Q)=\int_{NQ-\gamma Q}^{(N+1)Q-\gamma Q}\frac{\alpha}{x^2+\beta}dx\\
   &=\frac{\alpha}{\sqrt{\beta}}\left[\arctan\left(\frac{(N+1)Q-\gamma Q}{\sqrt{\beta}}\right)-\arctan\left(\frac{NQ-\gamma Q}{\sqrt{\beta}}\right)\right].
\end{split}
\end{equation}
The entropy of quantizated coefficients can be formulated by~\cite{li2009laplace},
\begin{equation}
\begin{split}
\label{3.6}
   H(Q) &=\sum_{N=-{L_{max}}}^{L_{max}}-P_N(Q)\log_2P_N(Q) \\
        &=-P_0(Q)\log_2P_0(Q)+2\cdot\sum_{N=1}^{L_{max}}-P_N(Q)\log_2P_N(Q).
\end{split}
\end{equation}
Herein, the $H(Q)$ is a monotonically decreasing function with \(Q\), as shown in Fig.~\ref{fig:relationshipQH}.

\begin{figure}[t]
\centering
\includegraphics[scale=0.5]{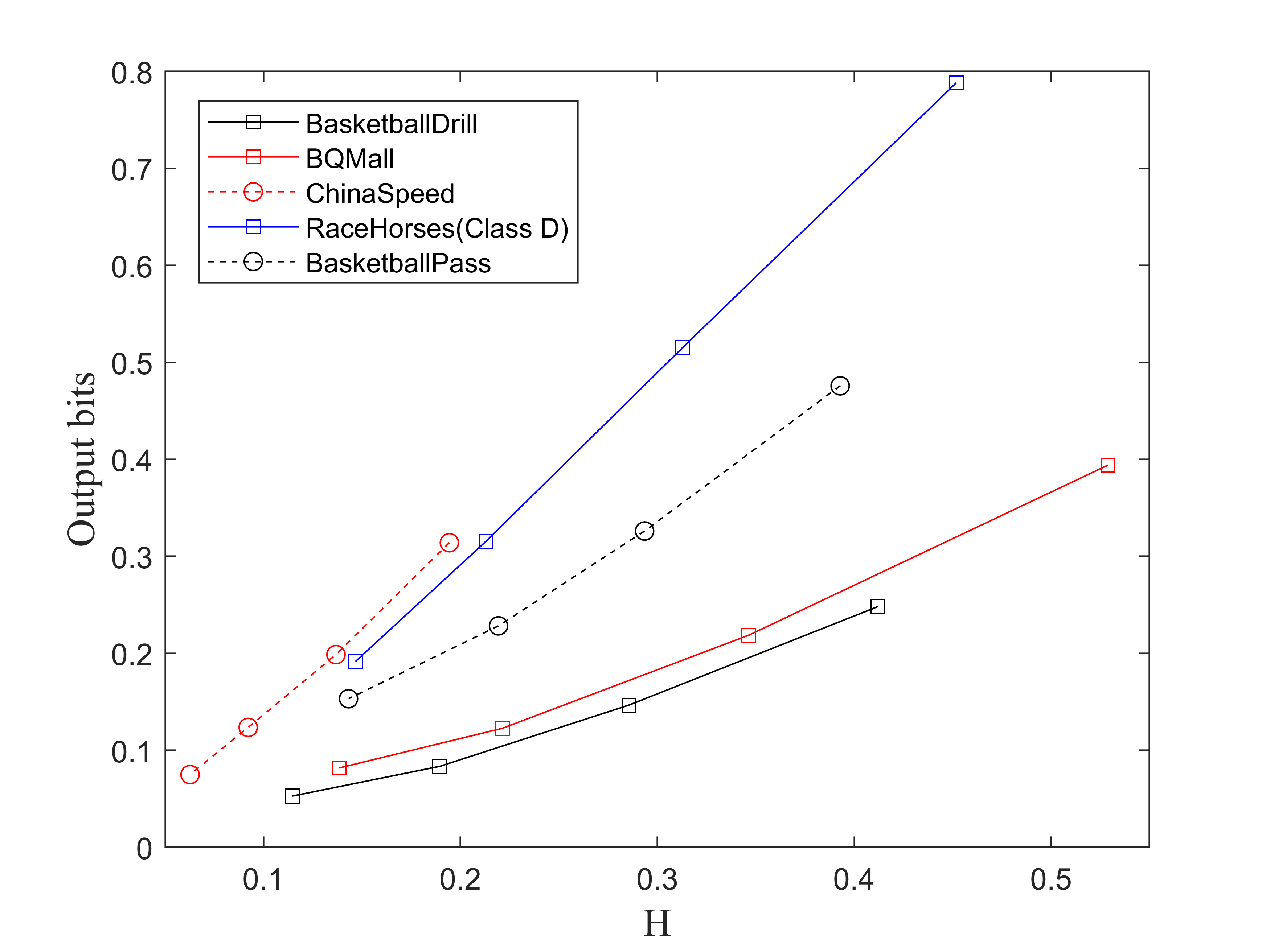}
\caption{Illustration of the relationship between estimated entropy of residuals and coding bits (per-pixel). 
}
\label{fig:relationshipRH}
\end{figure}

Subsequently, by performing the actual entropy coding, we exemplify the relationship between the estimated entropy and actual coding bits of five test sequences, as shown in Fig.~\ref{fig:relationshipRH}. In particular, the coding information of the 16-th frame is extracted from these sequences, where the $QP$s are set to 23, 28, 33 and 38. An approximate linear relationship between the estimated entropy and actual number of output coding bits can be observed.
As such, the coding bits of the current frame can be estimated as,
\begin{align}
\label{R_Qrelation_0}
    \hat{R}(Q)&=\phi \cdot H(Q) + \psi ,
\end{align}
where the slope \(\phi\) is characterized by the relationship between the actual coding bits of residuals and entropy, and the intercept \(\psi\) is determined by the header bits of the current frame. However, as these parameters cannot be obtained before encoding the current frame, we adopt a strategy to infer them from the previously coded frame at the same level. In particular, 
\begin{align}
  \phi = \frac{R_p-r_p^h}{H(Q_p)}
  \quad \text{and} \quad \psi=r_p^h,
\end{align}
where $R_p$ denotes the actual output bits (per-pixel) of the previously coded frame. Analogously, $Q_p$ represents the corresponding quantization step size of the previous frame, and $r_p^h$ denotes the header bits of previous frame which is also evaluated in terms of bits per pixel. Given the target rate, the corresponding QP is obtained by locating the corresponding \(Q\) that leads to the minimization between the frame-level target bits $\hat{R}_i$ and estimated encoding bits $\hat{R}(Q)$.

\subsection{$D$-$Q$ Model}
Given the quantization step size $Q$, the quantization distortions in terms of mean square error (MSE) can be estimated as follows,
\begin{equation}
\begin{split}
\label{3.8}
    &D_0(Q)=\sum_{n=\lfloor-Q+\gamma Q\rfloor}^{\lfloor(Q-\gamma Q)\rfloor} n^2\cdot\frac{\alpha}{n^2+\beta},\\
    &D_N(Q)=\sum_{n=\lfloor(N\cdot Q-\gamma Q)\rfloor}^{\lfloor(N+1)Q-\gamma Q\rfloor}(n-N\cdot Q)^2\frac{\alpha}{n^2+\beta}, \\
    &D_{-N}(Q)=\sum_{n=\lfloor(-(N+1)Q+\gamma Q)\rfloor}^{\lfloor-N\cdot Q+\gamma Q\rfloor}(n+N\cdot Q)^2\frac{\alpha}{n^2+\beta}.\\
\end{split}
\end{equation}
For simplicity, Eqn.~(\ref{3.8}) can be approximated by calculating definite integral as follows,
\begin{equation}
\begin{split}
\label{3.9}
     D_0(Q)&=\int_{-(Q-\gamma Q)}^{Q-\gamma Q}x^2\cdot\frac{\alpha}{x^2+\beta}dx \\
    & =2\alpha(Q-\gamma Q)-2\alpha\sqrt{\beta}\arctan(\frac{Q-\gamma Q}{\sqrt{\beta}}),\\
     D_N(Q)&=\int_{NQ-\gamma Q}^{(N+1)Q-\gamma Q}(x-NQ)^2\cdot\frac{\alpha}{x^2+\beta}dx \\
    & =\alpha Q+\Psi_1(Q,N)-\Psi_2(Q,N),
\end{split}
\end{equation}
where
\begin{align}
    \Psi_1(Q,N)=&\frac{\alpha N^2Q^2-\alpha \beta}{\sqrt{\beta}} \nonumber \\
    \cdot &\arctan\left(\frac{Q\sqrt{\beta}}{\beta+(NQ-\gamma Q)(NQ+Q-\gamma Q)}\right),\\
    \Psi_2(Q,N)=&\alpha NQ\cdot\ln\left(\frac{(NQ+Q-\gamma Q)^2+\beta}{(NQ-\gamma Q)^2+\beta}\right).
\end{align}
As such, the total distortion can be formulated as follows,
\begin{equation}
\begin{split}
\label{3.10}
    &D(Q) =D_0(Q)+2\cdot\sum_{N=1}^{L_{max}} D_N(Q).
\end{split}
\end{equation}
\begin{figure}[t]
\centering
\includegraphics[scale=0.5]{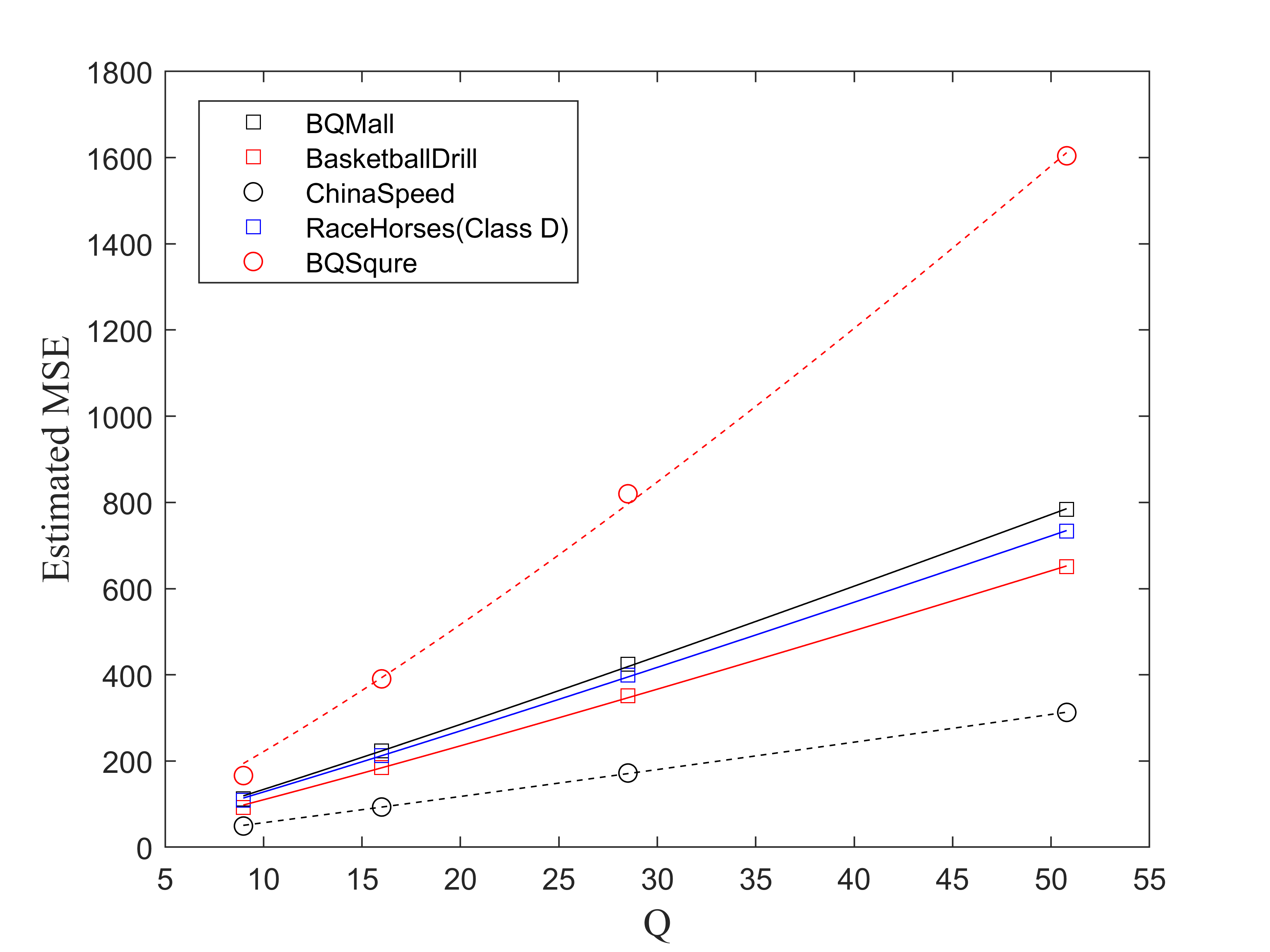}
\caption{Relationship between $Q$ and estimated distortion $D$.}\label{fig:relationshipQD}
\end{figure}
In Fig.~\ref{fig:relationshipQD}, the relationship between \(Q\) and \(D(Q)\) is shown, which further verifies that  \(D(Q)\) is a monotonically increasing function of \(Q\).

In real encoding scenarios, to compensate the influences of loop filters, dependent quantization as well as the SKIP-coded blocks, the distortion $\hat{D}$ of the current frame can be estimated with the adaptation of the distortion information of the previously coded frame as follows,
\begin{equation}
\begin{split}
\label{D_Qrelation}
    \hat{D}(Q)&=\frac{D_p^{ns}}{D(Q_p)}\cdot D(Q)\cdot(1-P_p^s)+ P_p^s\cdot D_p^s 
\end{split}
\end{equation}
Herein, $Q_p$ and $D_p^{ns}$ represent quantization step size and the distortion for non-SKIP coded blocks of the previously coded frame. For SKIP-coded blocks, we assume the associated coding bits are zero and the incurring distortion as $D_p^s$. $P_p^s$ is the ratio of SKIP-coded blocks measured in terms of the pixels within the previously coded frame.

\section{The Proposed Rate Control}
In this section, the rate control scheme is presented based on the proposed $\hat{R}$-$Q$ and $\hat{D}$-$Q$ models. First, the bit allocation scheme regarding the GOP-level and frame-level is elaborately designed wherein the inter-frame dependencies are comprehensively investigated. Subsequently, we present the derivation of coding parameters given the target bit-rate. Finally, the initialization and clipping strategy of coding parameters are discussed.
\subsection{Bit Allocation}
\subsubsection{GOP Level Bit Allocation}
Given the target bit-rate of a sequence $R^{t}_{seq}$, the ideal output bits for each GOP are derived as follows,
\begin{equation}
\begin{split}
\label{4.1}
    &R_{gop}^{t}=\frac{R^{t}_{seq}}{N_{GOP}}.
\end{split}
\end{equation}
Here, $N_{GOP}$ denotes the number of GOPs in a sequence. 
Since the actual output bits may deviate from the target bits because of diversified video contents, we employ a sliding window~\cite{li2014lambda} to flatten the output bits.
In particular, the mechanism behind the sliding window is that if the encoded frames consume more bits, the target bits for the following GOPs within the sliding window will be decreased accordingly and vice versa. As such, the target bits for the $g$-th GOP can be derived as,
\begin{equation}
\begin{split}
\label{4.2}
    R^{t}_{gop_g}=R_{gop}^{t}-\frac{R_{cost}-R_{gop}^{t}\cdot N_{coded}}{N_{SW}},
\end{split}
\end{equation}
where $R_{cost}$ denotes the cost of bits for all encoded frames, and $N_{SW}$ represents the size of the slide window. $N_{coded}$ is the number of frames that have already been encoded.

\begin{figure}[t]
\centering
\subfigure[]{   
\includegraphics[scale=0.8]{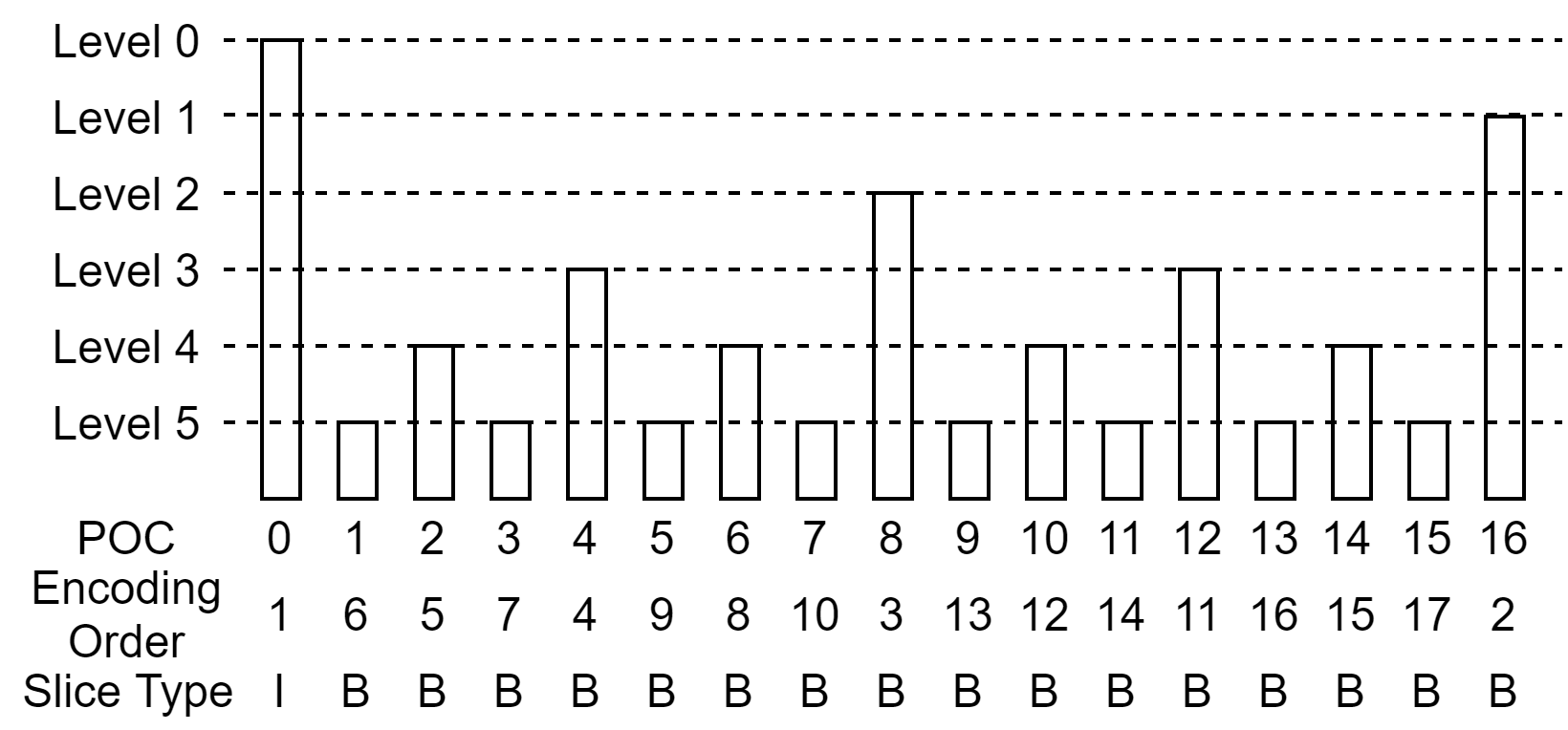}}\\
\subfigure[]{   
\includegraphics[scale=0.8]{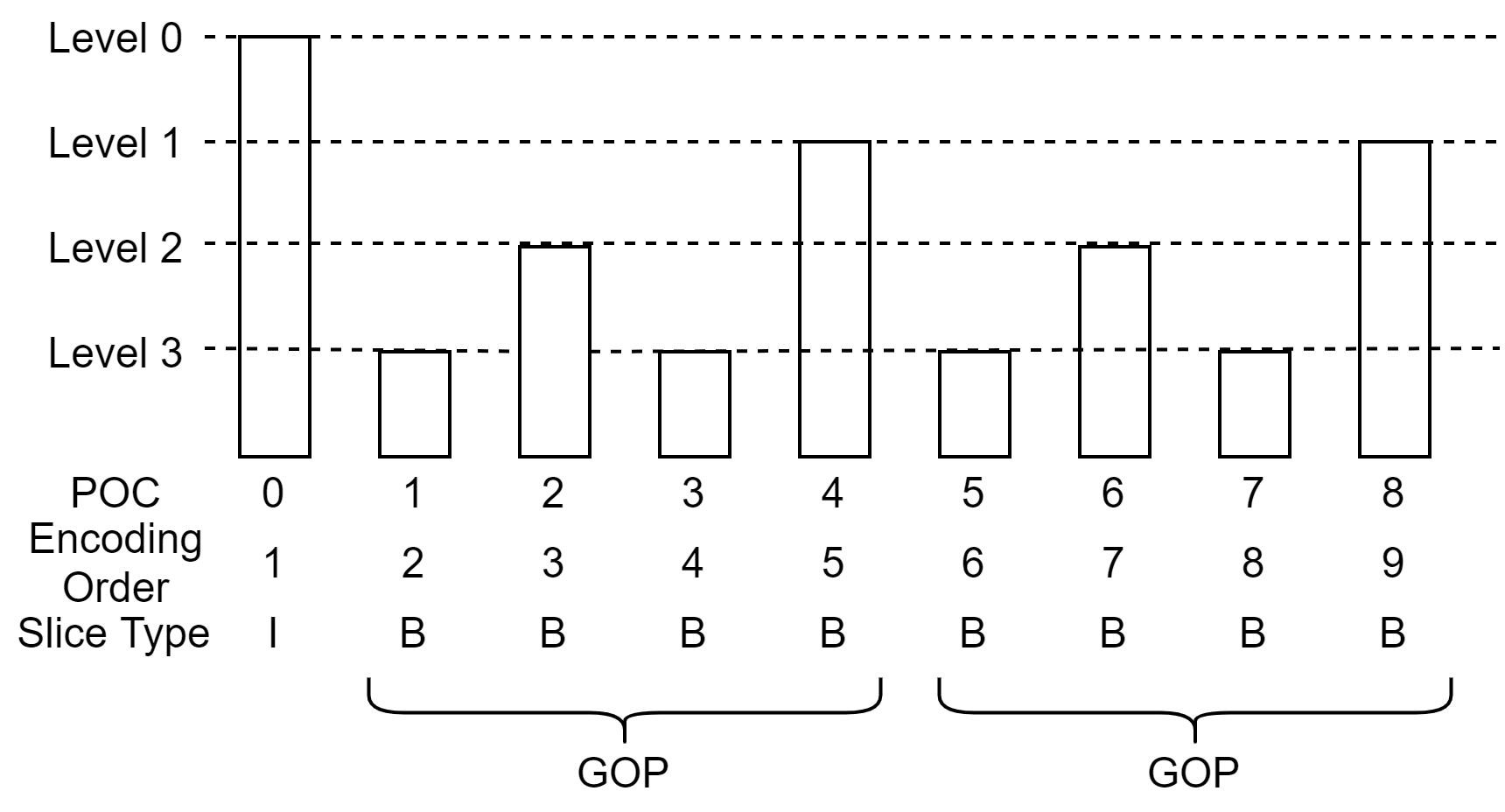}}
\caption{Frame structures of RA and LD in VVC~\cite{10026807777}. (a) RA structure with GOP size equaling to 16. (b) LD structure with GOP size equaling to 4.}
\label{fig:GOP_structure}
\end{figure}

\subsubsection{Frame Level Bit Allocation}
Two typical GOP structures in VVC are shown in Fig.~\ref{fig:GOP_structure} illustrating the hierarchical referencing relationship.
Regarding the bit allocation at the frame level, the inter-frame dependencies are fully considered. More specifically,
due to inter prediction in P and B-frames, there exists quality dependencies between the reference frame and the current to-be-coded frame. One widely accepted view is that the frames in lower temporal layers (\textit{i.e.} level 0), which may have more significant influences to the subsequent coding frames, are eligible to be assigned with more coding bits. In turn, less coding bits are assigned to the frames in higher temporal layers.
As such, the importance of different frames can be discriminated according to the referencing relationship as well as video content. In the literature, how reference frames affect the to-be-coded frame~\cite{hu2011rate,wang2013rate,li2016lambda,he2017adaptive} have been intensively investigated, {where a linear relationship regarding the coding distortions of reference frame and current one is noticed. Moreover, the existing schemes are also typically developed based on the strong assumption that the coding bits of the reference frame have negligible influence on the output bits of the current frame.} 
Considering that new coding tools have been adopted in VVC, in this paper, we revisit this problem based on new statistics collected in VTM-3.0~\cite{L1001-v9}, in an effort to explore the rate and distortion characteristics in the reference frame and the current to-be-coded frame. 

\begin{figure}[t]
\centering
\subfigure[]{   
\includegraphics[scale=0.28]{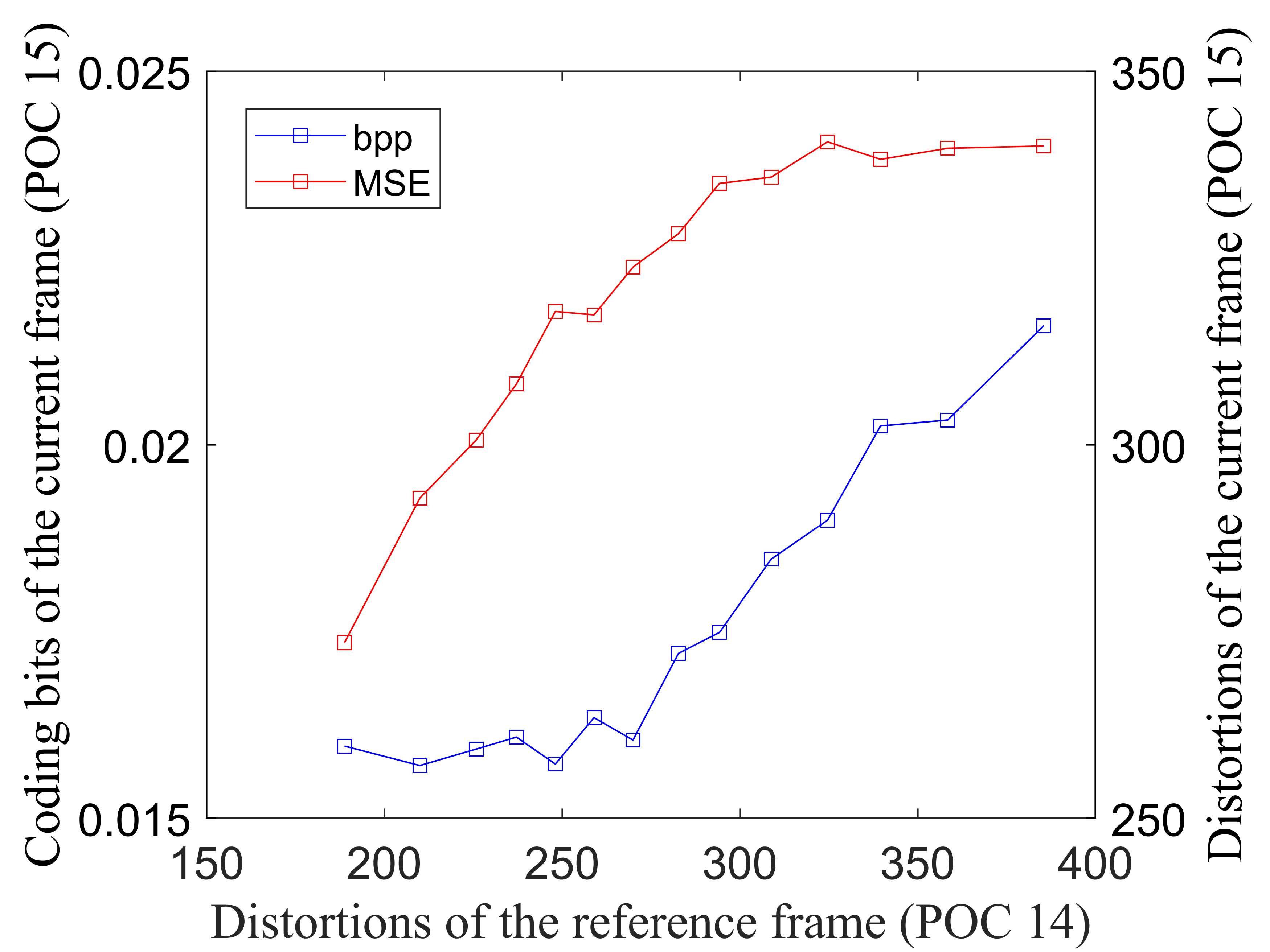}}
\subfigure[]{   
\includegraphics[scale=0.28]{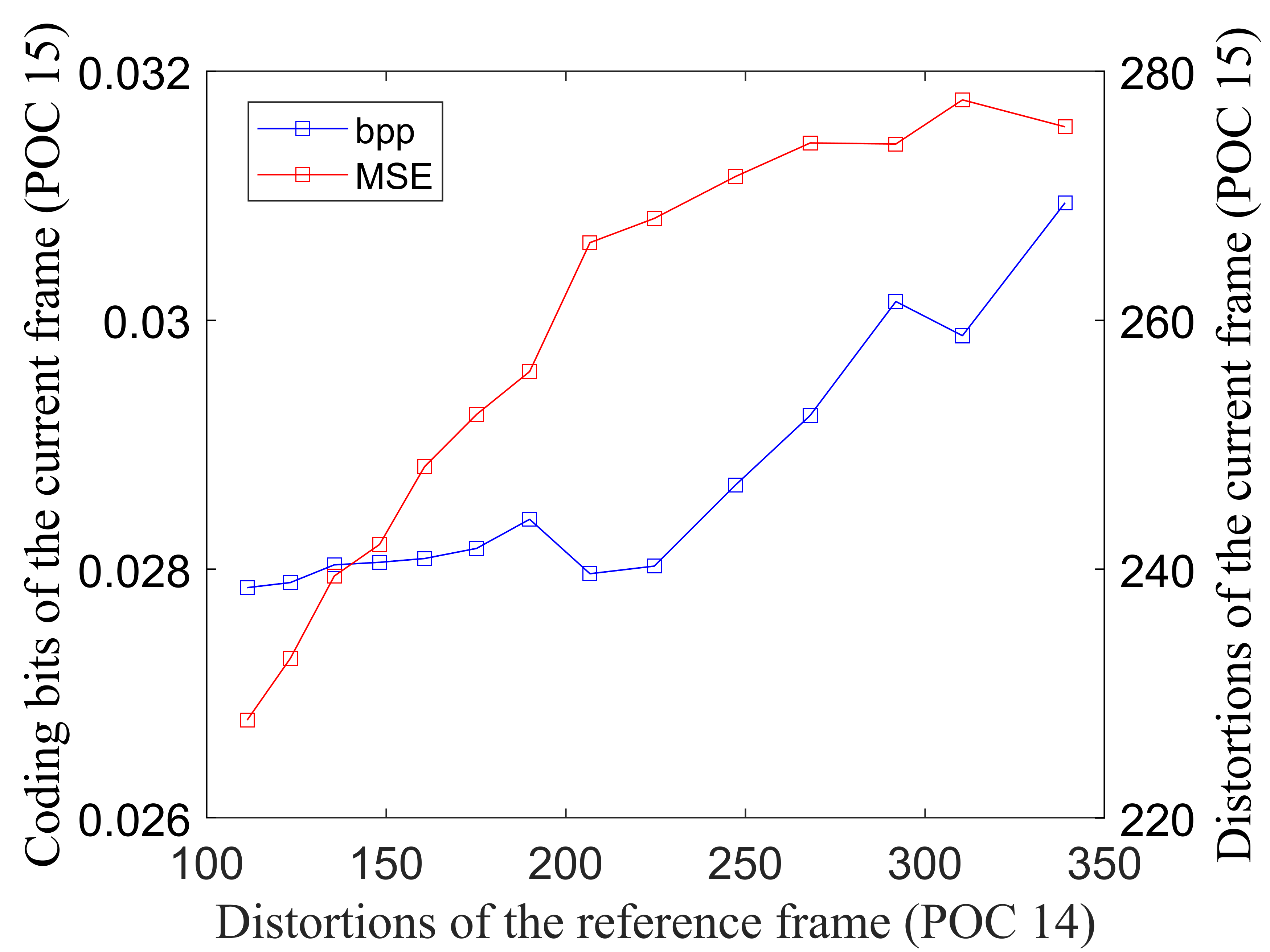}}\\
\subfigure[]{   
\includegraphics[scale=0.28]{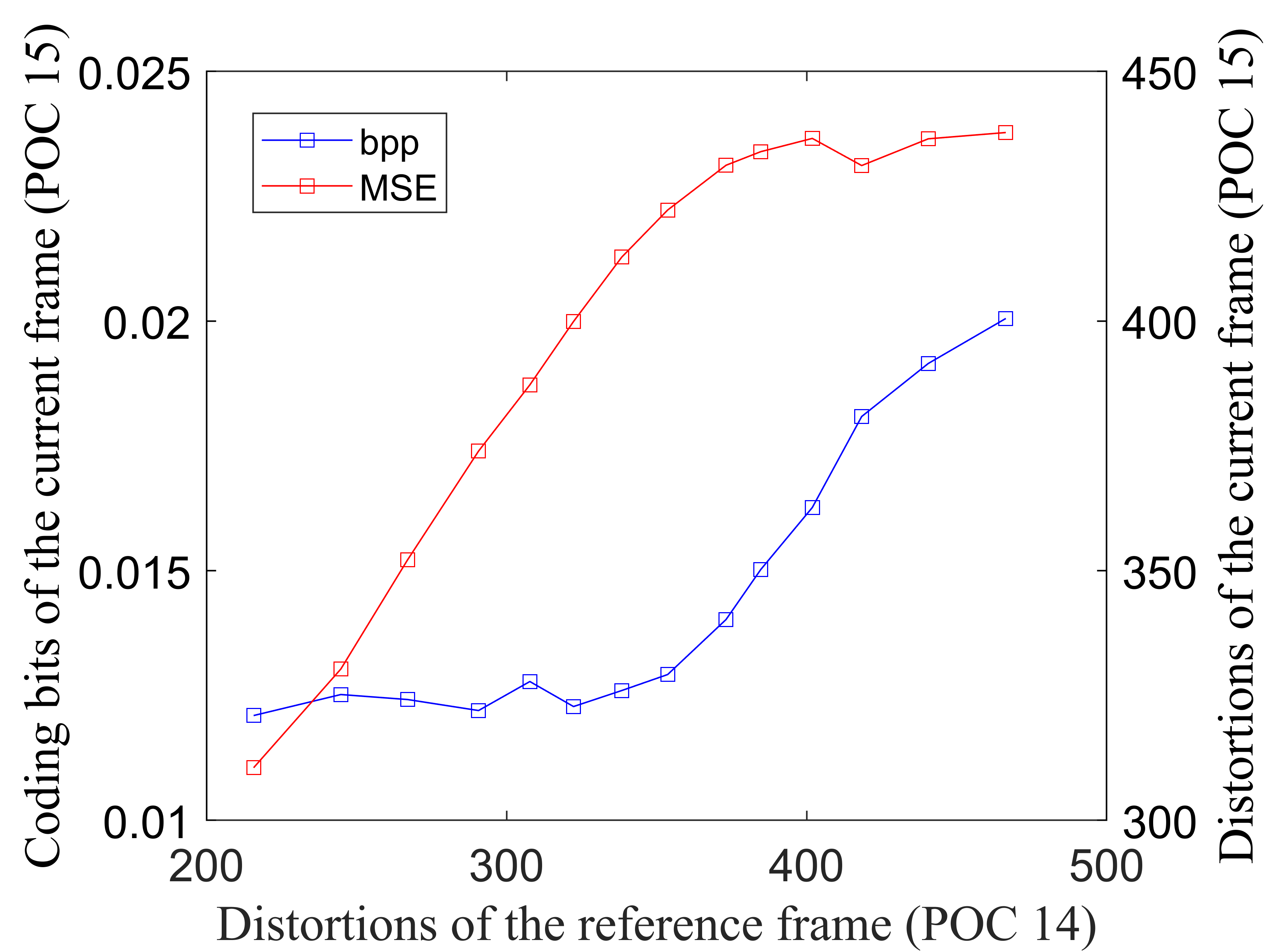}}
\subfigure[]{   
\includegraphics[scale=0.28]{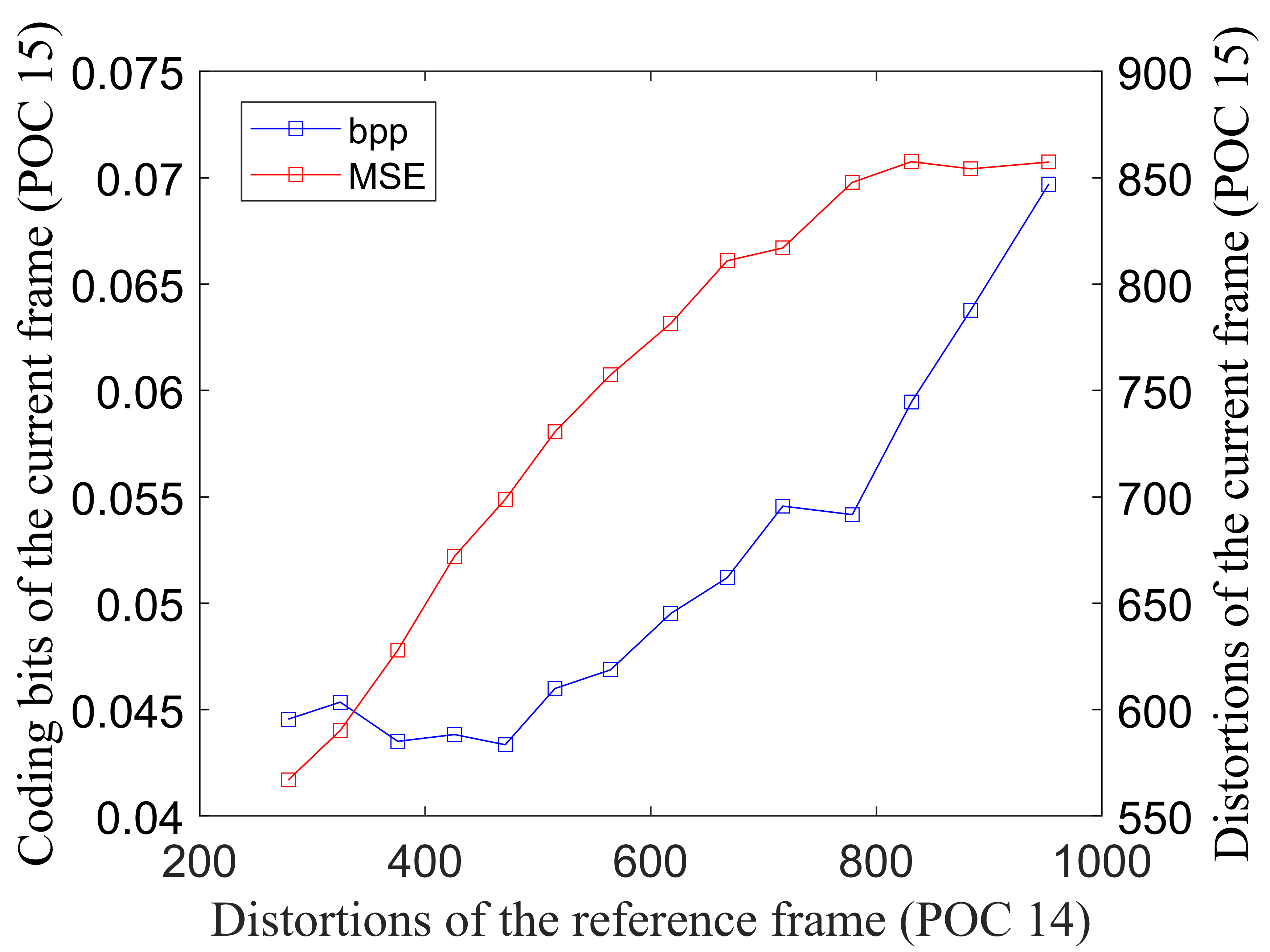}}
\caption{Illustration of the rate and distortion dependencies between the reference frame and the current coded frame. The x-axis denotes the MSE of the reference frame. The left and the right y-axis represent the output coding bits (per-pixel) and the MSE of the current coded frame, respectively. (a) ``BasketballDrill'' (b) ``ChinaSpeed'' (c) ``BQMall'' (d) ``RaceHorses''. 
}\label{fig:dependency1}
\end{figure}

As illustrated in Fig.~\ref{fig:dependency1}, the {quality of the }reference frame influences both the distortions and the coding bits of the current frame.
More specifically, four sequences are involved in the investigation under LDB configuration. For the current to-be-coded frame,
the associated $QP$ is fixed to 40. Meanwhile, the $QP$ of the reference frame
varies from 30 to 43, in an effort to generate references with different quality levels. We plot the corresponding output bits and distortions of the current frame with varying quality of the reference frame in Fig.~\ref{fig:dependency1}. We can observe that the distortions and coding bits of the current frame increase with the increment of the distortions in the reference frame. Moreover, it is interesting to see that the distortion increment of the reference frame leads to a linear augmented distortion of the current frame, along with a flat trend when the 
distortion of the reference frame reaches a certain level. The output coding bits (per-pixel) of the current frame varies smoothly when the reference frame is of high quality and increases sharply when the reference frame is severely distorted. These observations are in contrast to the existing models where only the distortion of the current to-be-coded frame is influenced by the quality of the reference frame. 

\begin{figure}[t]
\centering
\subfigure[]{   
\includegraphics[scale=0.28]{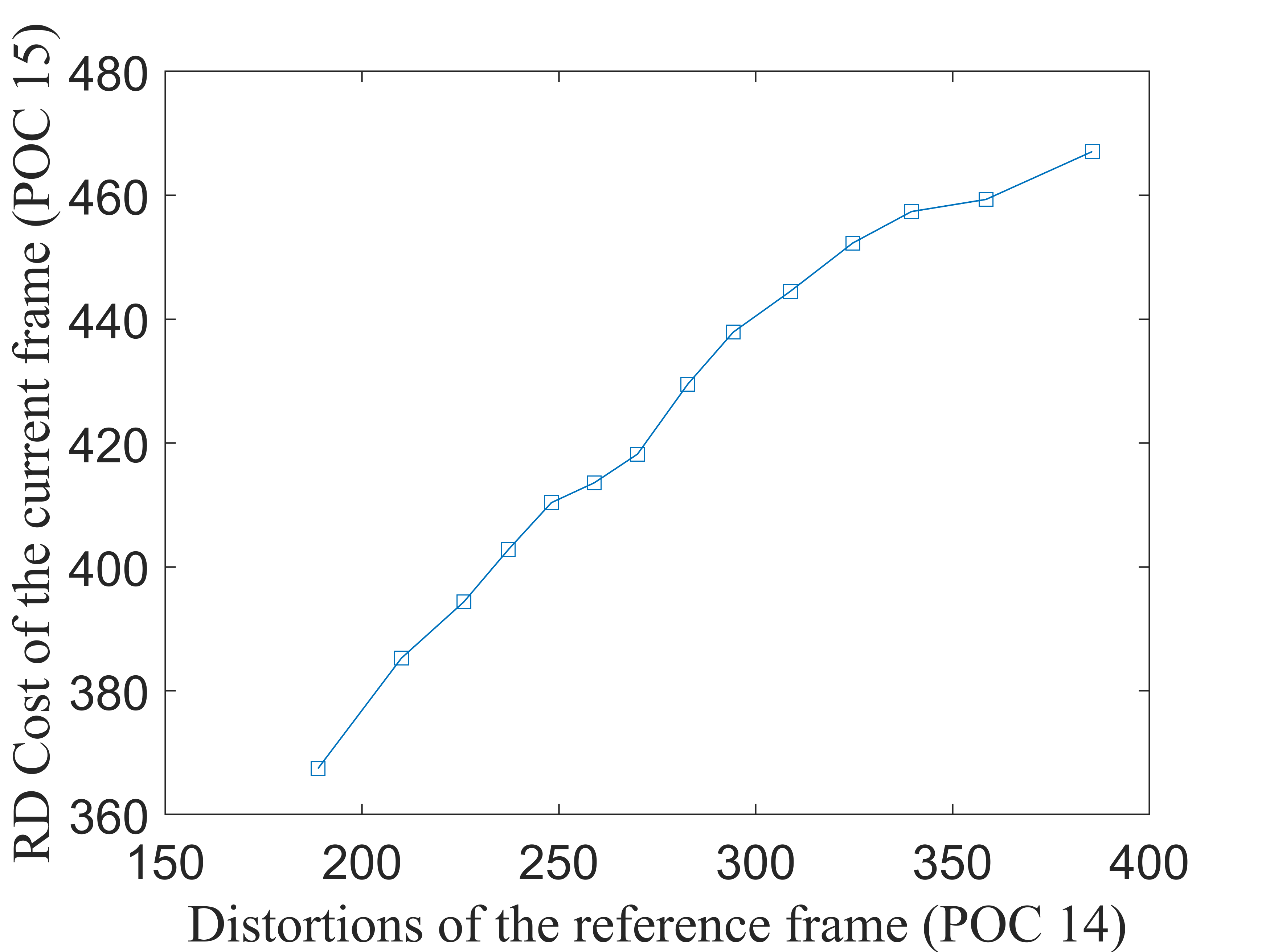}}
\subfigure[]{   
\includegraphics[scale=0.28]{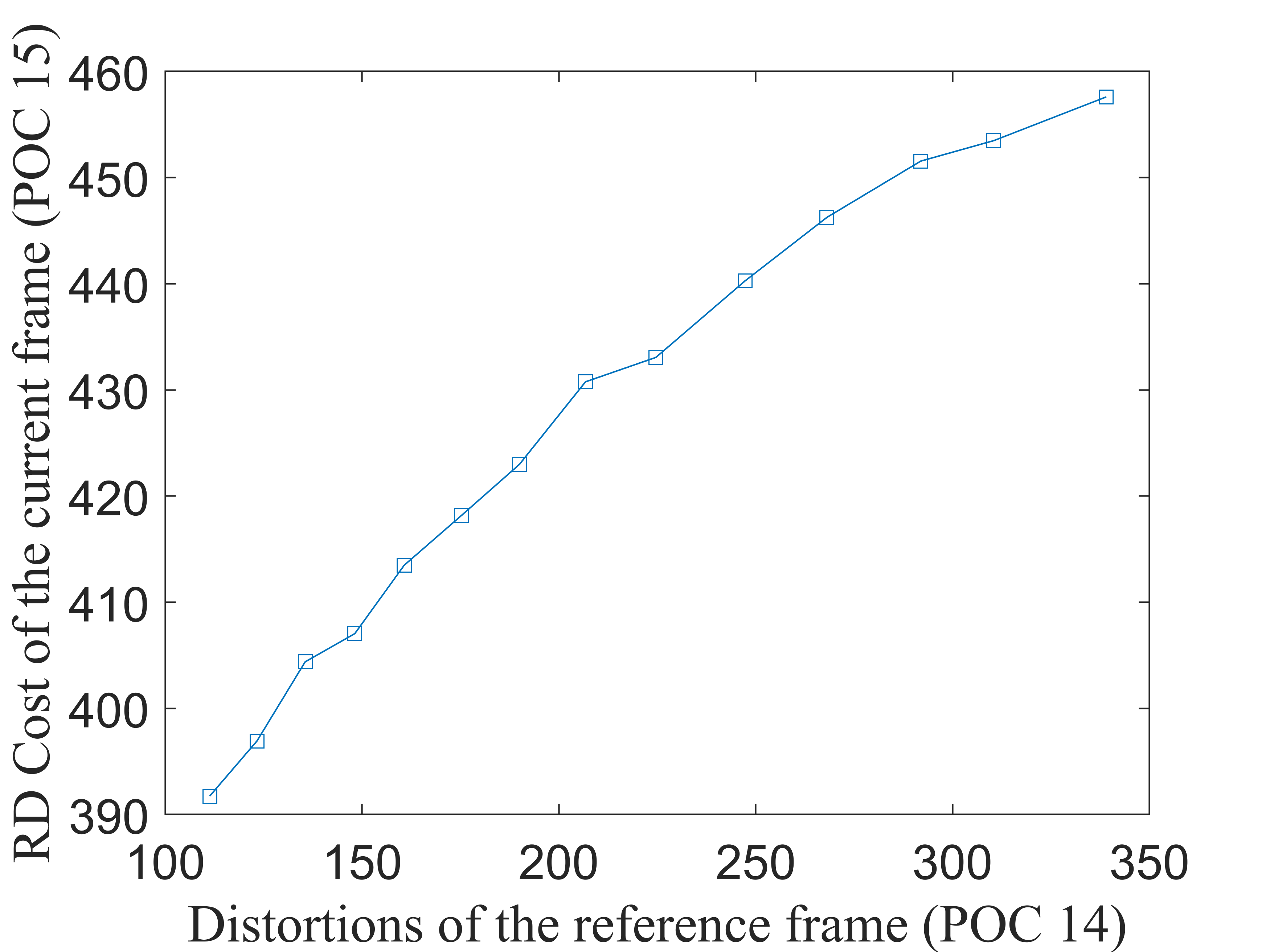}}\\
\subfigure[]{   
\includegraphics[scale=0.28]{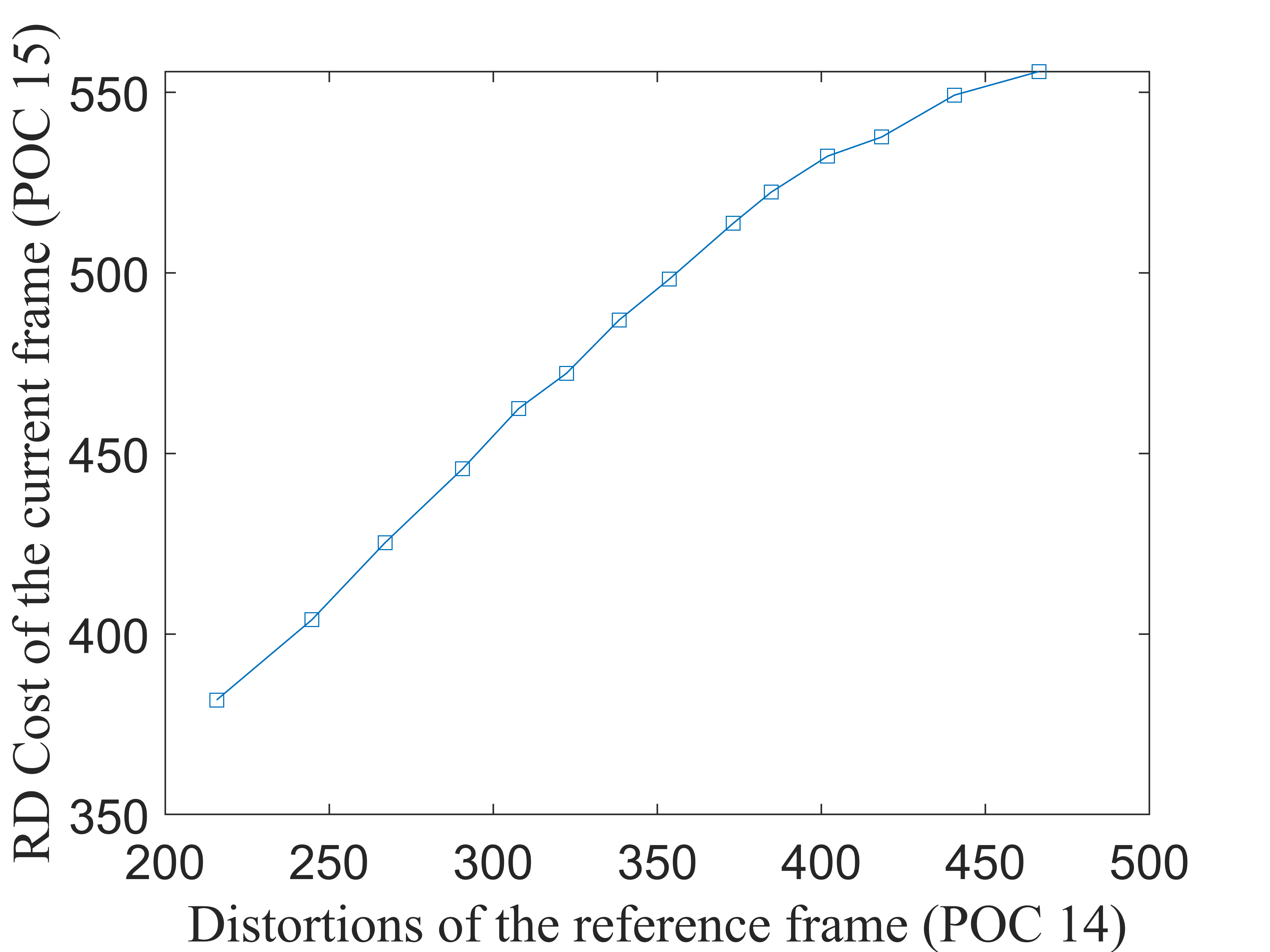}}
\subfigure[]{   
\includegraphics[scale=0.28]{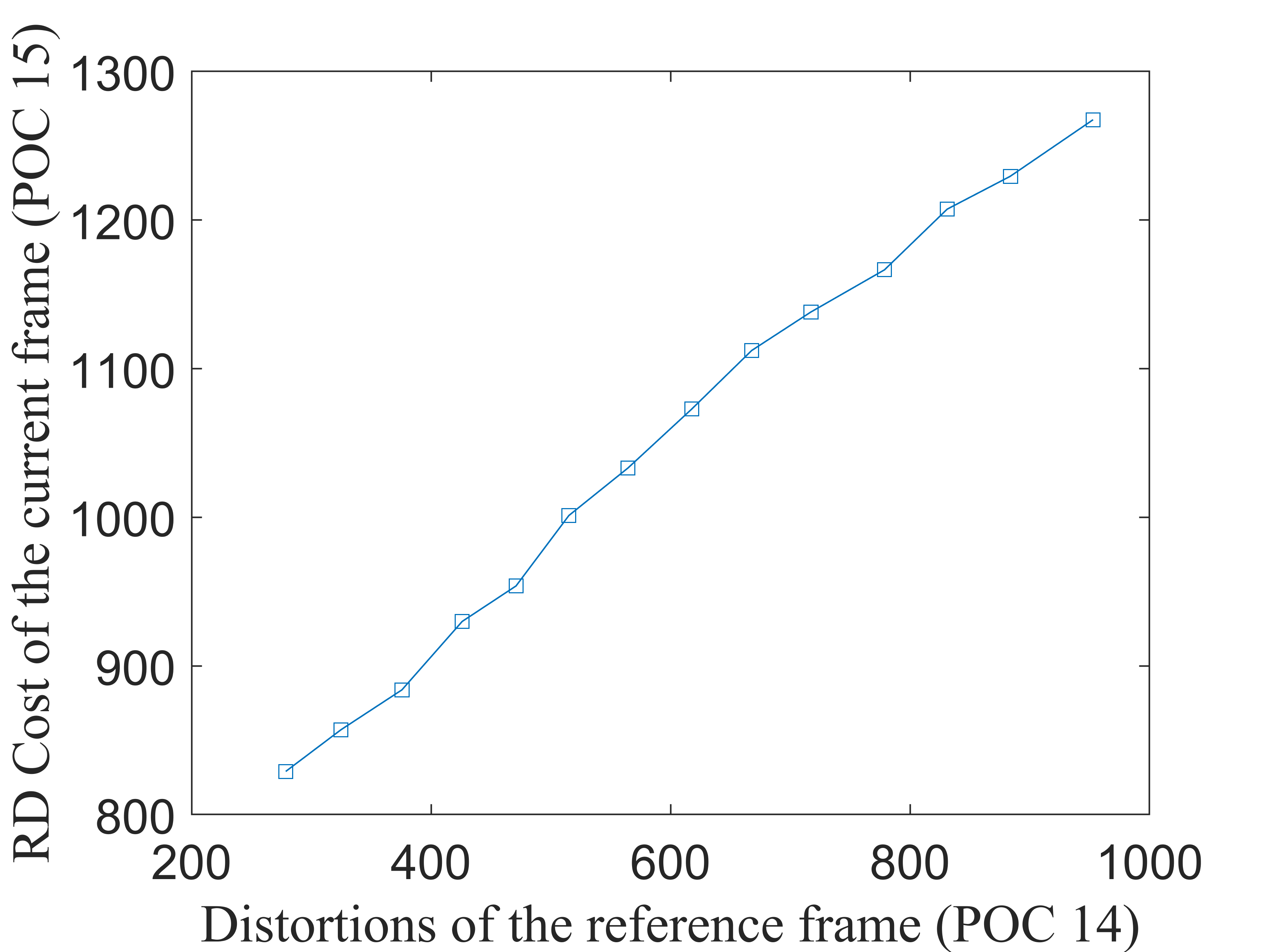}}
\caption{Illustration of the relationship between the distortion of the reference frame and the RD cost of the current coded frame. (a) ``BasketballDrill'' (b) ``ChinaSpeed'' (c) ``BQMall'' (d) ``RaceHorses''.
}\label{fig:dependency2}
\end{figure}

Considering the influences of both distortion and coding bits, there exists an approximately linear relationship between the distortion of reference frame and the RD cost of the current frame, as shown in Fig. \ref{fig:dependency2}. 
As such, we define the dependency factor $\pi^i_j$ between reference frame $j$ and encoding frame $i$ as follows,
\begin{equation}
\begin{split}
\label{4.3}
    \pi^i_j=\frac{dJ_i}{dD_j},
\end{split}
\end{equation}
where $J_i$ denotes the RD cost of the encoding frame and $D_j$ represents the distortion of the reference frame. 

Typically, the total RD cost of a GOP is formulated as the sum of the RD cost of each frame. 
Generally speaking, the distortion and coding bits of each frame  characterized by the Eqn.~(\ref{R_Qrelation_0}) and Eqn.~(\ref{D_Qrelation}) are highly dependent on the distribution parameter estimated, and in practice due to the chicken-egg-dilemma 
we could only use the statistics of the previous frame sharing the same level to estimate RD cost of the to-be-encoded frame. 
However, due to the influence of the reference frame quality, the straightforward estimation of the distribution parameters may lead to inaccurate modelling of the RD cost. In particular, we assume the distortion of the reference frame that serves for the previous frame as $D_p^{j_p}$, where $j_p$ belongs to previous frame's reference list. As such, the actual quality of the reference frame deviates from $D_p^{j_p}$, leading to the biased RD-cost estimated. 
To compensate for the RD cost difference introduced by quality fluctuation of the reference frames, the RD cost of each frame is formulated as the sum of internal RD cost $J_{in}^i$, external RD cost $J_{ex}^i$ and constant RD cost $J_{i}^c$. In particular, $J_{in}^i$ is derived based on Eqn.~(\ref{R_Qrelation_0}) and Eqn.(~\ref{D_Qrelation}), and $J_{ex}^i$ is incurred by difference between $D_p^{j_p}$ and distortion values of the reference frames within current GOP, such that it can be represented as  $\{\hat{D_j}(Q_j)-D_p^{j_p}\}$. $J_{i}^c$ is brought by difference between $D_p^{j_p}$ and distortion of the reference frames outside 
the current GOP. In other words, it could be regarded as a constant value.
As such, supposing there are $N_f$ frames in current GOP, the total RD cost of a GOP can be written as,
\begin{align}
\label{4.6}
    J_{total}=&\sum_{i=1}^{N_f} (J_{in}^i+J_{ex}^i+ J_{i}^c)\nonumber\\
    =&\sum_{i=1}^{N_f} [(\hat{D_i}(Q_i)+\lambda_{GOP} \hat{R_i}(Q_i))\nonumber\\&+\sum_j J_{ex}^i(\hat{D_j}(Q_j)-D_p^{j_p})+ J_{i}^c],
\end{align}
where $j$ is the index of reference list regarding the encoding frame. $Q_i$ and $Q_j$ denote the quantization step sizes of the current frame and reference frame $j$, respectively. As proved in Appendix, Eqn.~(\ref{4.6}) can be written as,
\begin{align}
\label{4.7}
J_{total}=&\sum_{i=1}^{N_f} [(\hat{D_i}(Q_i)+\lambda_{GOP} \hat{R_i}(Q_i))\nonumber\\
&+\sum_k J_{ex}^k(\hat{D_i}(Q_i)-D_p^{i_p})+J_{i}^c]\nonumber\\
=&\sum_{i=1}^{N_f} J_i(Q_i).
\end{align}
Herein $k$ is the index of frame list which uses current frame $i$ as a reference and $D_p^{i_p}$ is distortion for the previous frame of frame $i$. $J_i$ is sum of the internal RD cost of a frame and its influence on other frames. In order to minimize the total RD cost of a GOP of frames $J_{total}$, we need to find the optimal $Q_i$ for individual frame. Considering $J_i$ is a function of $Q_i$ and $Q_i$ is an independent parameter, $J_i$ of frame $i$ is independent from other frames' $QP$s. To minimize $J_{total}$ which is the sum of $J_i$, we need to minimize each $J_i$ individually.
As such, we compute the partial derivation of $J_i$ with respect to $Q_i$, which is set equaling to 0 as follows,
\begin{align}
\label{4.8}
{\frac{\partial J_i}{\partial Q_i}} &= \left(\frac{\partial \hat{D_i}(Q_i)}{\partial Q_i}+\lambda_{GOP} \frac{\partial \hat{R_i}(Q_i)}{\partial Q_i}\right)\nonumber\\
    &+\sum_k \frac{\partial J_{ex}^k(\hat{D_i}(Q_i)-D_p^{i_p})}{\partial \hat{D_i}(Q_i)}\cdot\frac{\partial \hat{D_i}(Q_i)}{\partial Q_i}=0 .
\end{align}
According to the former analyses that there exists an approximated linear relationship between the distortion of the reference frame and the RD cost of the current encoding frame, by integrating Eqn.~(\ref{4.3}) into Eqn.~(\ref{4.8}), we can obtain,
\begin{equation}
\begin{split}
\label{4.9}
    {\frac{\partial J_i}{\partial Q_i}}=& \left(\frac{\partial \hat{D_i}(Q_i)}{\partial Q_i}+\lambda_{GOP} \frac{\partial \hat{R_i}(Q_i)}{\partial Q_i}\right)+\frac{\partial \hat{D_i}(Q_i)}{\partial Q_i}\cdot\sum_k \pi_i^k\\
    =& \left(\kappa_i \cdot\frac{\partial \hat{D_i}(Q_i)}{\partial Q_i}+\lambda_{GOP} \frac{\partial \hat{R_i}(Q_i)}{\partial Q_i}\right)=0,\\
\end{split}
\end{equation}
where $\kappa_i$ is the influence factor,
\begin{equation}
\begin{split}
\label{4.10}
    \kappa_i=1+\sum_k \pi_i^k.
\end{split}
\end{equation}
The influence factor reveals the importance of a frame. More specifically, frames with higher $\kappa_i$ have greater impact on other frames, 
deserving to be assigned with more coding bits.
In this optimization problem, the whole GOP shares the same $\lambda_{GOP}$, 
\begin{equation}
\begin{split}
    \lambda_{GOP}=-\frac{\kappa_i \cdot\frac{\partial \hat{D_i}(Q_i)}{\partial Q_i}}{\frac{\partial \hat{R_i}(Q_i)}{\partial Q_i}}
\end{split}
\end{equation}
Here, we need to obtain derivatives of Eqn.~(\ref{R_Qrelation_0}) and Eqn.~(\ref{D_Qrelation}). However, the complex nature of  Eqn.~(\ref{3.6}) and Eqn.~(\ref{3.9}) 
makes it difficult for us to obtain analytical $R-Q$ and $D-Q$ relationships. In~\cite{kamaci2005frame}, the hyperbolic function is used to model Cauchy distribution based $R-Q$ and $D-Q$ relationships. Inspired by this method, we obtain different combinations of $\{Q, \hat{R}(Q)\}$ and $\{Q, \hat{D}(Q)\}$ and model them with hyperbolic function. Derivatives of the two fitting models are used to approximate derivatives of Eqn.~(\ref{R_Qrelation_0}) and Eqn.~(\ref{D_Qrelation}), which 
are denoted as $\hat{R}^{'}(Q)$ and $\hat{D}^{'}(Q)$.
For frame $i$, the associated QP candidates are from $QP_p^i-3$ to $QP_p^i+3$, where $QP_p^i$ denotes the QP used to encode previous frame. 
Given the derivatives of $R-Q$ and $D-Q$, we utilize Algorithm~\ref{algorithm:bits} to search allocated bits to each frame to ensure the optimal RD performance as well as the satisfaction of the bit-rate budget. 

\begin{algorithm}[t] 
\caption{Optimal bit allocation.}
\label{algorithm:bits}
\begin{algorithmic}[]
\Require Frame $i$'s QP candidate list $QP_i^1, QP_i^2, \dots, QP_i^7$ and the corresponding RD model within the current GOP.
\Ensure Target bit-rate $R_i^t$ for frame $i$ in current GOP.
    \For {$v\text{ from 1 to 7}$}
    \begin{itemize}
        \item[] \textbf{Step 1:} Supposing level 1 frame is the $j$-th frame within GOP and its $v$-th candidate QP is $QP^{v}_j$ of which the corresponding quantization step size is $Q^{v}_j$. Slopes for R-Q and D-Q curve at $Q^{v}_j$ are $\hat{R}_{j}^{'}(Q^{v}_j)$ and $\hat{D}_{j}^{'}(Q^{v}_j)$ respectively. By denoting $\hat{R}_{j}^{'}(Q^{v}_j)$ and $\hat{D}_{j}^{'}(Q^{v}_j)$ as $\hat{R}_{j_{v}}^{'}$ and $\hat{D}_{j_{v}}^{'}$, we can define $\lambda_{GOP}^{v}$ as,
        \begin{align}
        \label{Equation_init}
        \lambda_{GOP}^{v}=-\frac{\kappa_j \cdot \hat{D}_{j_{v}}^{'}}{\hat{R}_{j_{v}}^{'}}.
        \end{align}
        \item[] \textbf{Step 2:}  
        Select optimal QP for frame $i$ from its QP candidate list: $QP_{i}^1, QP_{i}^2, \dots, QP_{i}^7$. 
        
        \begin{align}
            \label{Equation_test}
            \lambda_{u}=-\frac{\kappa_i \cdot \hat{D}_{i_{u}}^{'}}{\hat{R}_{i_{u}}^{'}}, u\text{ from 1 to 7}.
        \end{align}
        We can obtain:
        \begin{align}
            u_{i{v}}=\min_{u}{|\lambda_{u}-\lambda_{GOP}^{v}|}.
        \end{align}
        $QP_i^{u_{i{v}}}$ is selected as the optimal QP of frame $i$ and stored in a QP list.
        
        \item[] \textbf{Step 3:} The ${v}$-th QP list can be written as: $QP^{u_{1{v}}}_1, QP^{u_{2{v}}}_2, \dots, QP^{u_{{N_f}{v}}}_{N_f}$. 
    \end{itemize}
    \EndFor
    \begin{itemize}
        \item[] \textbf{Step 4:} Obtain target bits for each frame. Supposing corresponding quantization step size of the $v$-th QP list is $Q^{u_{1{v}}}_1, Q^{u_{2{v}}}_2, \dots, Q^{u_{{N_f}{v}}}_{N_f}$. By combining Eqn.~(\ref{R_Qrelation_0}), the order of optimal QP list ${v}_o$ is obtained as,
        \begin{align}
            {v}_{o}=\min_{{v}}{|(\sum_{i=1}^{{N_f}}\hat{R}_i(Q^{u_{i{v}}}_i)-R_{gop}^{t}|}, {v}\text{ from 1 to 7}.
        \end{align}
        Bits allocated to frame $i$ is given by,
        \begin{align}
            R_i^t=\hat{R}_i(Q^{u_{i{v}_o}}_i).
        \end{align}
    \end{itemize}
\end{algorithmic} 
\end{algorithm}

\subsection{Coding Parameters Derivation}
After obtaining the target bit-rate $R^t_i$, the coding parameters $\lambda_i$ and $QP_i$ can be derived according to Eqn.~(\ref{R_Qrelation_0}). 
Given the QP candidate list of frame $i$, the quantization step $Q_i$ can be calculated as,
\begin{align}
    \label{Equation_QP_selection}
    &Q_i=\min_{Q}{|\hat{R}_i(Q)-R^t_i|},\\ 
    &QP\in \{QP_p^i-3, QP_p^i-2, \dots, QP_p^i+3\}\nonumber
\end{align}
where $Q$ is the corresponding quantization step size of $QP$. 

Theoretically, $\lambda$ is the slope of RD curve, which can be derived as,
\begin{align}
\label{Q-lambda}
    \lambda^t_i(Q)=-\frac{\partial D_i}{\partial R_i}=-\frac{\frac{\partial (\hat{D}_i+D_e)}{\partial Q}}{\frac{\partial (\hat{R}_i+R_e)}{\partial Q}}=-\frac{\hat{D}^{'}_i(Q)}{\hat{R}^{'}_i(Q)},
\end{align}
where $D_e$ and $R_e$ denote difference of distortion and bit-rate incurred by the discrepance of reference frame quality which could be  
regarded as constant parameters.
Moreover, we collect the coding information of three previous frames to ensure a stable $Q-\lambda$ relationship. Let $\{Q_p^m ,\lambda_p^m\}$ denote the quantization step size and $\lambda$ of the $m$-th previous frame on the same level, the stability is given by,
\begin{align}
    \Gamma_m = \frac{\lambda_p^m}{\lambda^t_i(Q_p^m)},\  1\leq m \leq 3
\end{align}
More specifically, the value of $\Gamma_m$ closing to 1 indicates that the derived $Q-\lambda$ relationship from Eqn.~\eqref{Q-lambda} is stabilized. 
$\Gamma_m$ is further used to scale $\lambda^t_i(Q_i)$, such that $\lambda_i$ can be obtained as,
\begin{align}
    \lambda_i=\frac{\sum\limits_{m=1}^3\tau_m\cdot \Gamma_m}{\sum\limits_{m=1}^3 \tau_m}  \cdot \lambda^t_i(Q_i),
\end{align}
Here, $\tau_m$ is a predefined parameter of which the value is 5, 3, 1 for $m$ equaling to 1, 2, 3 respectively.
\subsection{Initial Value and Parameter Clip}

The proposed rate control scheme is applied on P and B slices. In practical implementation, the first frame of each level is coded with default rate control algorithm. For the first 32 frames, a fixed-ratio bit allocation scheme is applied to train stable coding parameters for adaptive bit allocation. Regarding bit allocation under RA structure, we assume that frames in the same temporal level share the identical influence factor $\kappa_i$. The explicit values of $\kappa_i$ are shown in Table \ref{tab:gamma_i_ra}. LD configuration involves simpler reference relationship and smaller GOP size, such that the influence factor is more sensitive to the coding bits. We define four sets of influence factor for each frame in LD configuration according to bit-per-pixel (bpp), as shown in Table \ref{tab:gamma_i_lowdelay}, where $I_{G}$ is an integer larger than zero. To cater the original GOP structure, we add extra restrictions to QP as illustrated in Table \ref{tab:clip_lowdelay} and Table \ref{tab:clip_ra}. The $QP^{(z)}_p$ indicates the QP of the previous encoded frame at $z$-th frame level.

\begin{table}[t]
  \centering
  \caption{Influence Factor for RA}
  \label{tab:gamma_i_ra}
    \begin{tabular}{cc}
    \toprule
    Frame Level&Influence Factor\\
    \midrule
    1&5.4082\\
    2&2.3958\\
    3&1.5933\\
    4&1.1566\\
    5&1\\
    \bottomrule
    \end{tabular}
\end{table}
\begin{table}[t]
  \centering
  \caption{Influence Factor for LDB}
  \label{tab:gamma_i_lowdelay}
    \begin{tabular}{ccccc}
    \toprule
    \multirow{2}{*}{}&
    \multicolumn{4}{c}{POC ID}\\
    &$4\cdot I_{G}-3$&$4\cdot I_{G}-2$&$4\cdot I_{G}-1$&$4\cdot I_{G}$\\
   \midrule
    0$<$bpp$\leq$0.05&1.587&1.7802&1.3781&5.1715\\
    0.05$<$bpp$\leq$0.1&1.4499&1.6675&1.3631&3.6495\\
    0.1$<$bpp$\leq$0.15&1.2432&1.409&1.1175&3.3994\\
    0.15$<$bpp$\leq$0.2&1.3633&1.5461&1.3363&2.6198\\
    \bottomrule
    \end{tabular}
\end{table}
\begin{table}[t]
  \centering
  \caption{QP Clips for LD Configuration}
  \label{tab:clip_lowdelay}
    \begin{tabular}{ccc}
    \toprule
    Frame Level&Lower Bound& Upper Bound\\
    \midrule
    3&$QP^{(1)}_p$&-\\
    2&$QP^{(1)}_p$&$QP^{(3)}_p$\\
    1&-&$QP^{(3)}_p-4$\\
    \bottomrule
    \end{tabular}
\end{table}
\begin{table}[t]
  \centering
  \caption{QP Clips for RA Configuration}
  \label{tab:clip_ra}
    \begin{tabular}{ccc}
    \toprule
    Frame Level&Lower Bound& Upper Bound\\
    \midrule
    5&$QP^{(1)}_p$&$QP^{(1)}_p+13$\\
    4&$QP^{(1)}_p$&$QP^{(1)}_p+13$\\
    3&$QP^{(1)}_p$&$QP^{(1)}_p+10$\\
    2&$QP^{(1)}_p$&$QP^{(1)}_p+6$\\
    1&$QP^{(5)}_p-11$&$QP^{(5)}_p-4$\\
    \bottomrule
    \end{tabular}
\end{table}

\begin{table}[t]
  \centering
  \caption{Characteristics of Test Sequences}
  \label{tab:characteristic}
    \begin{tabular}{ccccc}
    \toprule
    Class&Number of&Resolution&Frame&Bit\\
    &Sequences&&Rate&Depth\\
    \midrule
     A1&3&4K&60\&30&10\\
     A2&3&4K&60\&50&10\\
     B&5&1080p&60\&50&8\&10\\
     C&4&WVGA&60\&50\&30&8\\
     D&4&WQVGA&60\&50\&30&8\\
     E&3&720p&60&8\\
    \bottomrule
    \end{tabular}
\end{table}
\begin{figure}[htp]
\centering
\subfigure[]{   
\includegraphics[scale=0.080]{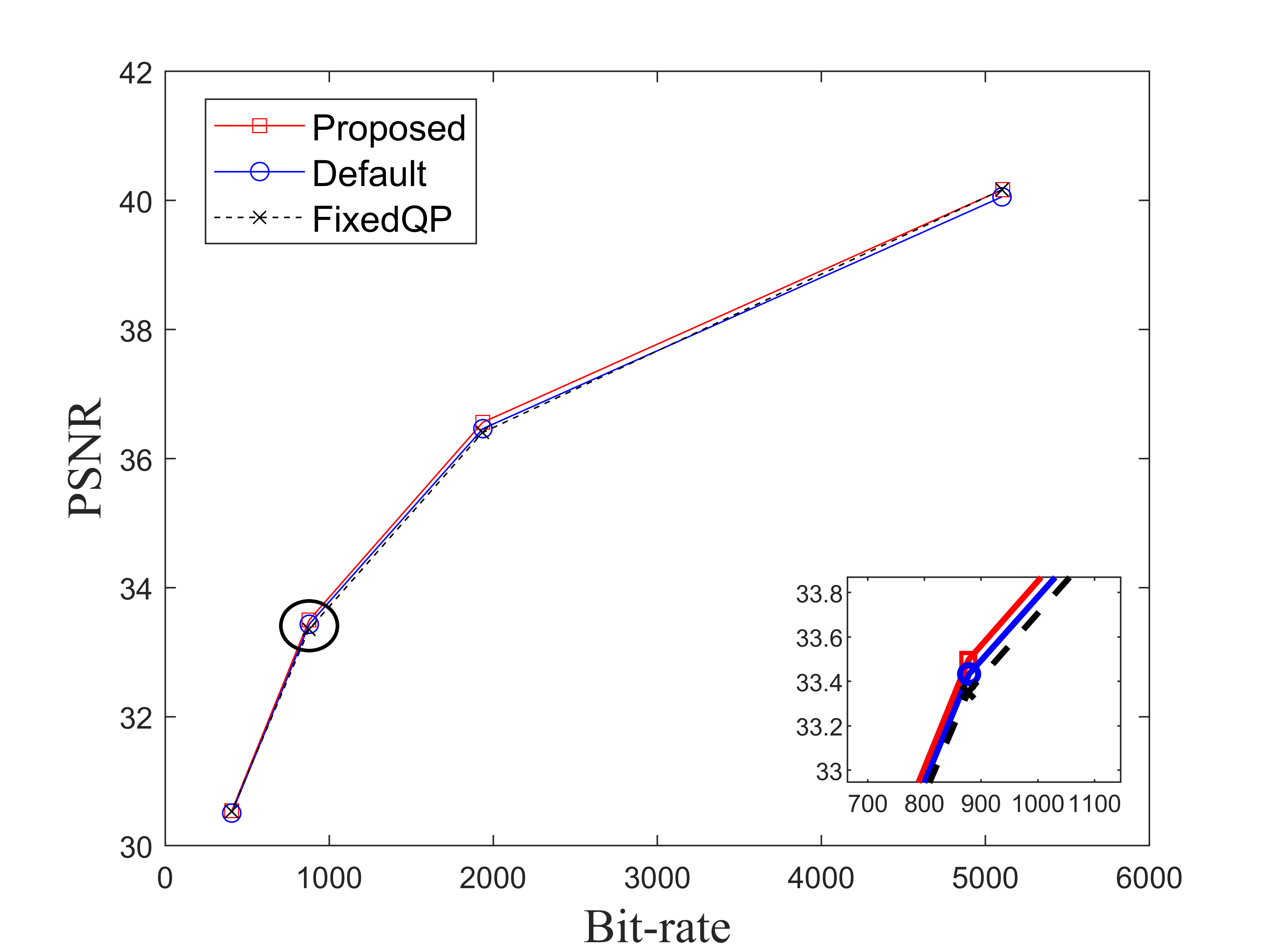}}\\
\subfigure[]{   
\includegraphics[scale=0.080]{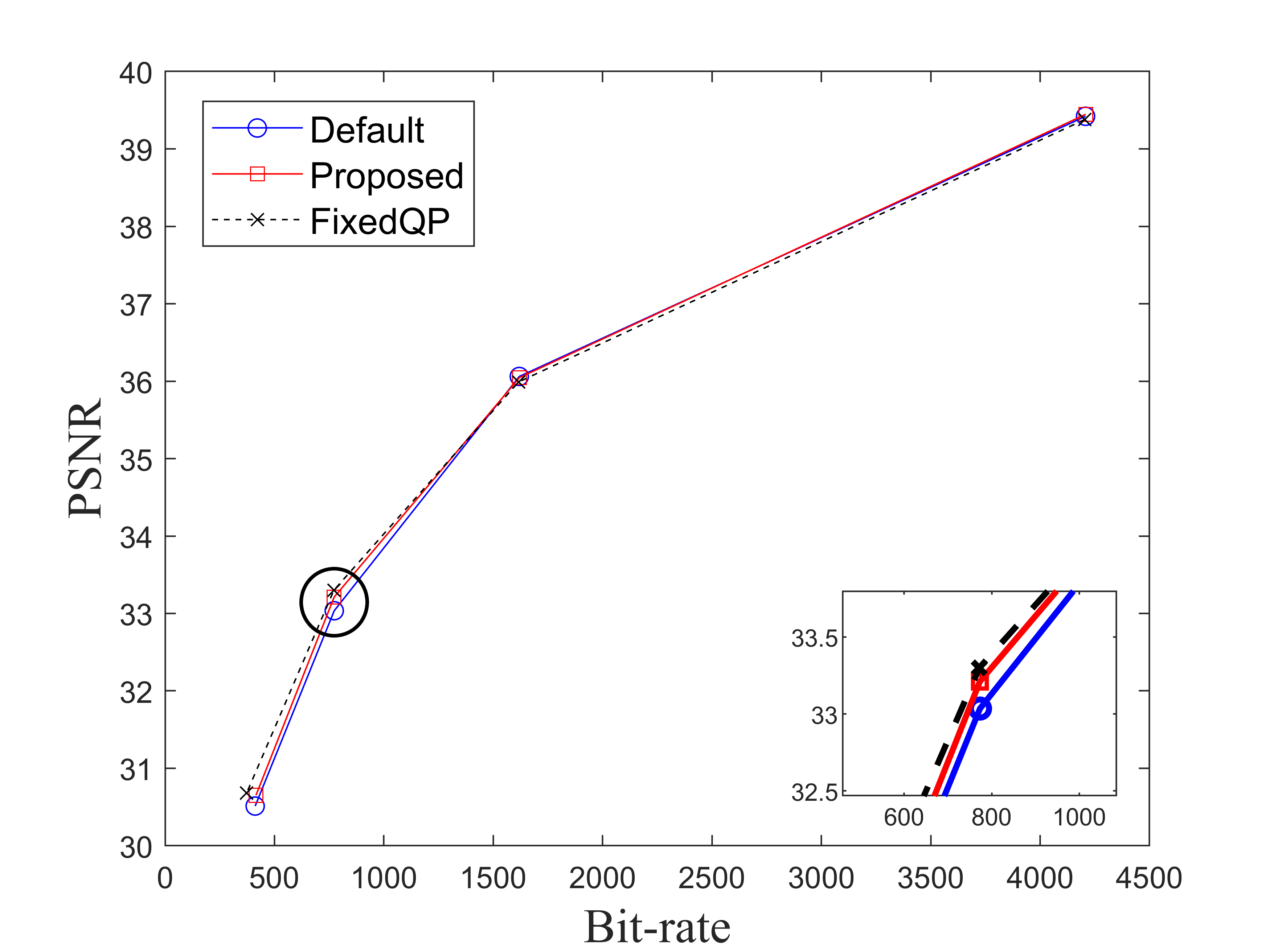}}
\caption{The RD curves of sequence ``RaceHorses'' (Class C) under LDB and RA configurations. (a) LDB configuration; (b) RA configuration.}
\label{fig:RD_compare2}
\end{figure}

\section{Experimental Results}
The proposed rate control algorithm is implemented on the VVC test model VTM-3.0~\cite{VTM3}. Extensive experiments are conducted to verify the effectiveness of the proposed method conforming to the common test conditions (CTCs)~\cite{L1010} under LDB (GOP size = 4) and RA (GOP size = 16) configurations. QPs are set to  22, 27, 32 and 37.
Details of recommended test sequences are summarized in Table~\ref{tab:characteristic}. Experiments are executed on a dual Intel Xeon CPU E5-2620 platform without parallelism.
We employ the original VTM-3.0 without rate control to encode test sequences following the CTCs, and regard the output bit-rate as the target bit-rate for rate control. The compression performance is measured with BD-Rate~\cite{BD-Rate} where
negative BD-Rate denotes the performance improvement.
In addition, the bit-rate error $BitErr$ is calculated to measure the rate control accuracy as follows,
\begin{equation}
\begin{split}
\label{5.1}
    BitErr=\frac{|R^{o}-R^{t}|}{R^{t}}\times100\%,
\end{split}
\end{equation}
where $R^{t}$ denotes the target bit-rate, and $R^{o}$ is the corresponding output bit-rate.

\subsection{Results and Analyses}

\begin{table}[t]
  \centering
  \caption{Illustration of the BD-Rate of the Proposed Rate Control Scheme on VTM-3.0 under LDB and RA Configurations}
  \label{tab:overall_BDRate}
    \begin{tabular}{ccccc}
    \toprule
    \multirow{3}{*}{}&
    \multicolumn{2}{c}{LDB}&\multicolumn{2}{c}{RA}\\
    &Fixed-QP&Default&Fixed-QP&Default\\
    &as anchor&as anchor&as anchor&as anchor\\
    \midrule
    Class A1&-&-&9.93\%&-3.03\%\\
    Class A2&-&-&3.49\%&-0.15\%\\
    Class B&-3.58\%&-1.24\%&3.76\%&-0.91\%\\
    Class C&-3.40\%&-0.48\%&1.58\%&-1.32\%\\
    Class D&-1.43\%&-0.08\%&3.30\%&-1.16\%\\
    Class E&-1.32\%&-1.43\%&-&-\\
    \midrule
    \textbf{Overall}&\textbf{-2.96\%}&\textbf{-1.03\%}&\textbf{4.36\%}&\textbf{-1.29\%}\\
    \midrule
    Enc. time&{125\%}&{123\%}&{121\%}&{118\%}\\
    \bottomrule
    \end{tabular}
\end{table}

\begin{figure}[htp]
\centering
\subfigure[]{   
\includegraphics[scale=0.45]{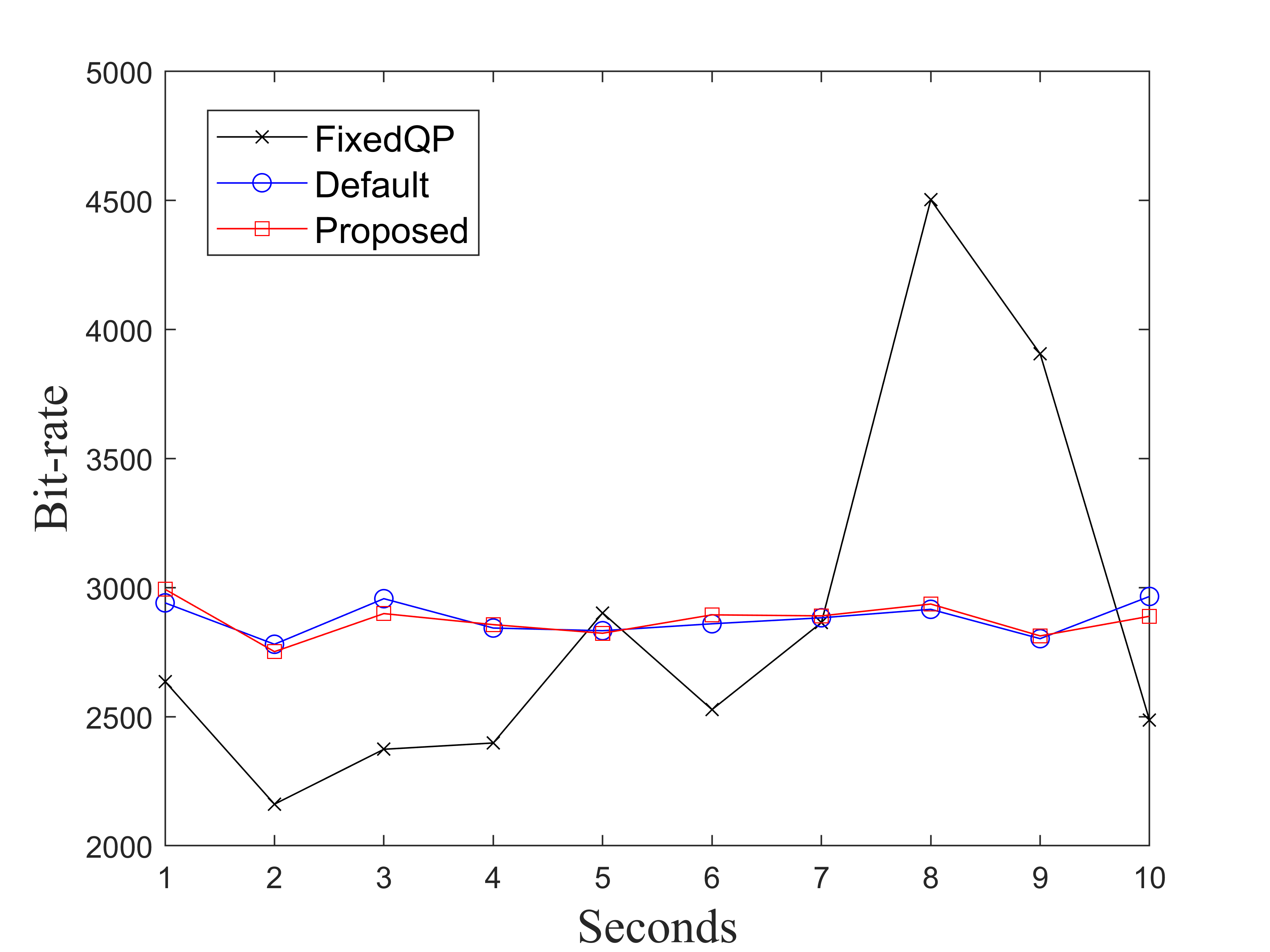}}\\
\subfigure[]{   
\includegraphics[scale=0.45]{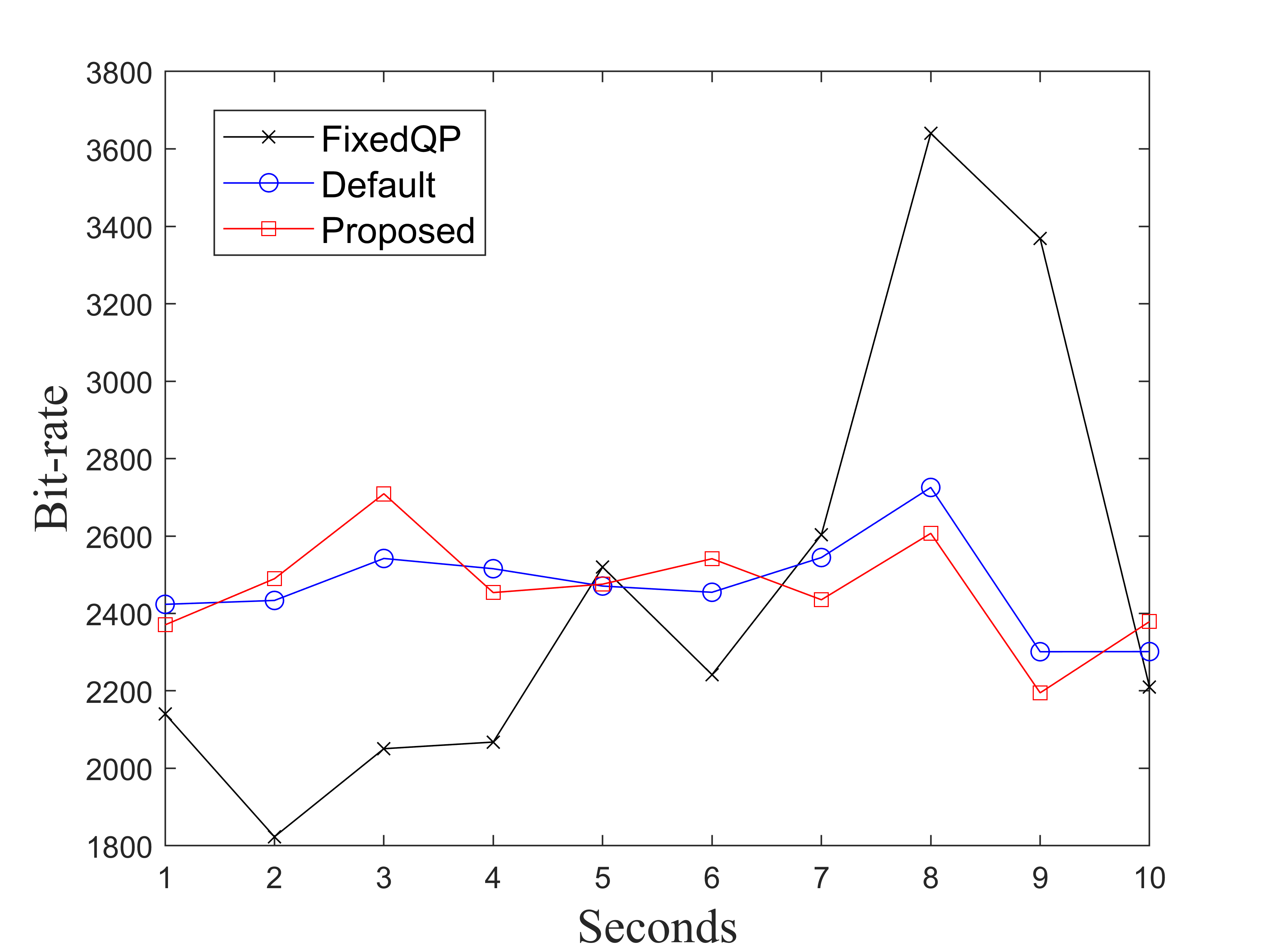}}
\caption{Illustration of the actual bits per-second for ``RitualDance''. (a) LDB configuration where the target bit-rate is set as 2876 kbps. (b) RA configuration where the target bit-rate is set as 2467 kbps.}
\label{fig:sequence}
\end{figure}

\begin{table}[t]
  \centering
  \caption{Illustration of the Average Bit-rate Error of the Default Rate Control and the Proposed Rate Control Schemes on VTM-3.0 under LDB and RA Configurations}
  \label{tab:bitrate_error}
    \begin{tabular}{ccc}
    \toprule
    &LDB&RA\\
    \midrule
    Proposed&0.3543\%&2.177\%\\
    Default&0.4158\%&2.635\%\\
    \bottomrule
    \end{tabular}
\end{table}

\begin{figure}[htp]
\centering
\subfigure[]{   
\includegraphics[scale=0.45]{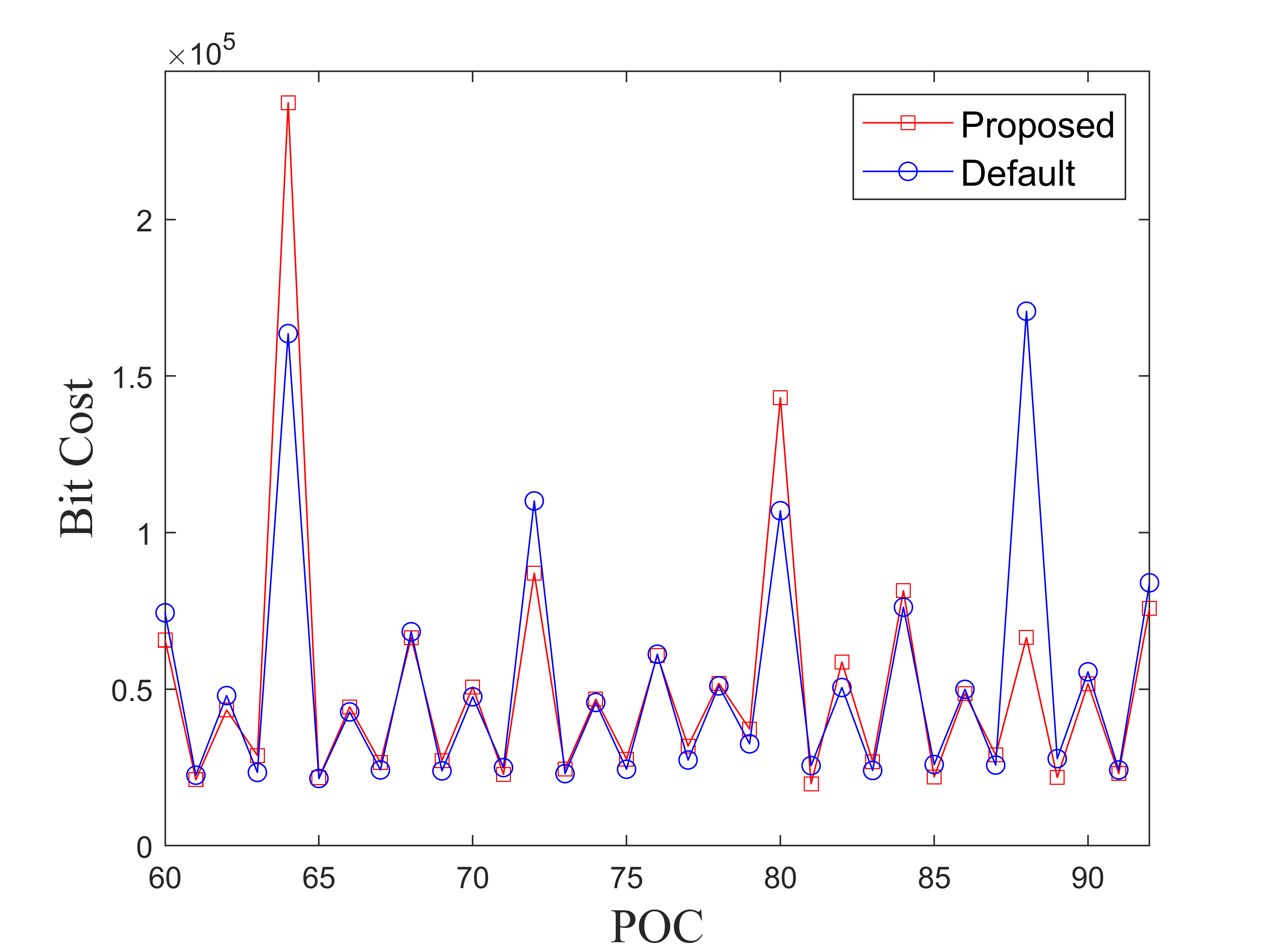}}\\
\subfigure[]{   
\includegraphics[scale=0.45]{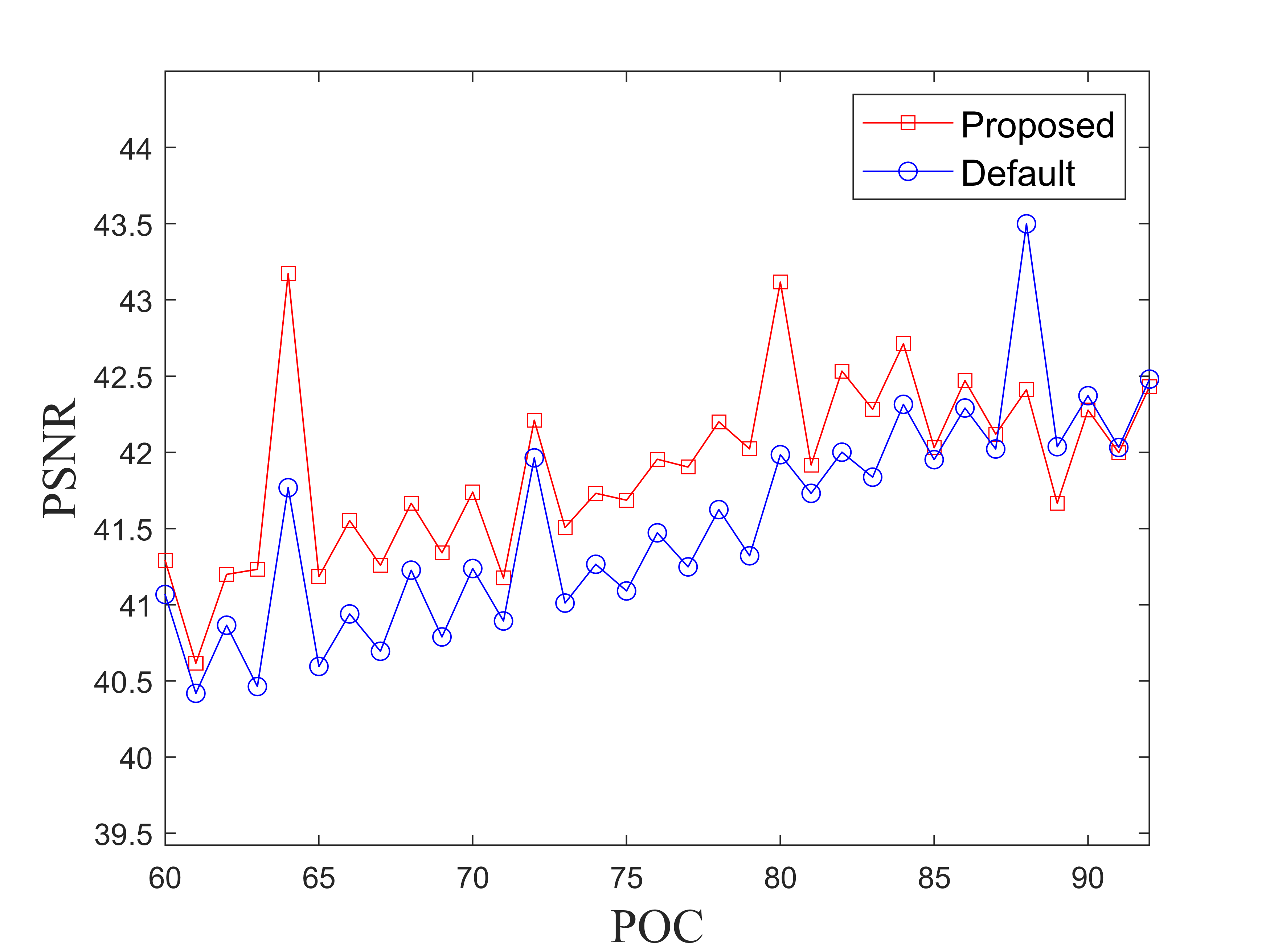}}
\caption{Illustration of the instant coding bits and PSNR of sequence ``BaksteballDrill'' with the proposed rate control scheme and the default rate control scheme from POC 60 to POC 92 under RA configuration. The target bit-rate is 2856 kbps. (a) Bit cost; (b) PSNR.
}

\label{fig:RD_compare1}
\end{figure}

\begin{table*}[htbp]
  \centering
  \caption{Experimental Results of ``BasketballDrive'' and ``BQMall'' under LDB Configuration}
  \label{tab:example_Lowdelay}
    \begin{tabular}{cccccccc}
    \toprule
     \multirow{2}{*}{Sequence}&\multirow{2}{*}{Target Bit-rate} &\multicolumn{3}{c}{Default Rate Control Algorithm}&\multicolumn{3}{c}{Proposed Rate Control Algorithm}\\
    &&Bit-rate&Y-PSNR&Bit-rate Error&Bit-rate&Y-PSNR&Bit-rate Error\\
   \midrule
    \multirow{4}{*}{BasketballDrive}&17189.78 & 17185.16 & 39.4888 & 0.027\% & 17167.99 & 39.5838 & 0.127\%  \\
    &5487.445 & 5490.41 & 37.6964 & 0.054\% & 5488.855 & 37.731 & 0.026\% \\
    &2605.99 & 2608.773 & 35.901 & 0.107\% & 2607.83 & 35.9127 & 0.071\% \\
    &1359.594 & 1361.379 & 33.948 & 0.131\% & 1359.434 & 33.9544 & 0.012\%  \\
    \midrule
    \multirow{4}{*}{BQMall}&3586.56 & 3590.396 & 40.3463 & 0.107\% & 3588.333 & 40.3821 & 0.049\%  \\
    &1565.79 & 1569.038 & 37.5914 & 0.207\% & 1568.903 & 37.6226 & 0.199\%  \\
    &771.25 & 773.8624 & 34.8017 & 0.338\% & 773.0672 & 34.8623 & 0.235\% \\
    &394.17 & 396.492 & 32.0209 & 0.590\% & 397.0336 & 32.1024 & 0.728\%  \\
    \bottomrule
    \end{tabular}
\end{table*}

\begin{table*}[htbp]
  \centering
  \caption{Experimental Results of ``BasketballDrive'' and ``BQMall'' under RA Configuration}
  \label{tab:example_Randomaccess}
    \begin{tabular}{cccccccc}
    \toprule
     \multirow{2}{*}{Sequence}&\multirow{2}{*}{Target Bit-rate} &\multicolumn{3}{c}{Default Rate Control Algorithm}&\multicolumn{3}{c}{Proposed Rate Control Algorithm}\\
    &&Bit-rate&Y-PSNR&Bit-rate Error&Bit-rate&Y-PSNR&Bit-rate Error\\
    \midrule
    \multirow{4}{*}{BasketballDrive}&14299.21 & 14303.8 & 39.4227 & 0.032\% & 14297.89 & 39.4029 & 0.009\% \\
    &4625.193 & 4628.515 & 37.7323 & 0.072\% & 4625.287 & 37.7437 & 0.002\% \\
    &2185.733 & 2203.797 & 35.9003 & 0.826\% & 2188.247 & 36.0319 & 0.115\% \\
    &1102.946 & 1248.451 & 33.8775 & 13.192\% & 1123.8808 & 34.0671 & 1.898\% \\
    \midrule
    \multirow{4}{*}{BQMall}&2894.882 & 2902.675 & 40.4221 & 0.269\% & 2896.334 & 40.4188 & 0.050\% \\
    &1293.245 & 1297.324 & 37.9194 & 0.315\% & 1294.236 & 37.9512 & 0.077\% \\
    &650.956 & 661.2928 & 35.36 & 1.588\% & 653.367 & 35.3822 & 0.370\% \\
    &334.54 & 352.3384 & 32.5389 & 5.320\% & 349.750 & 32.8307 & 4.546\% \\
    \bottomrule
    \end{tabular}
\end{table*}

Table~\ref{tab:overall_BDRate} shows the coding performance of proposed rate control algorithm under LDB and RA configurations. The original VTM-3.0 anchor without rate control (fixed-QP) and the default frame-level rate control algorithm in VTM-3.0 are respectively employed as the benchmark for comparison. As required by~\cite{L1010}, class D is excluded from the overall average.
In particular, compared with the default rate control algorithm, the proposed scheme brings $1.03\%$ and $1.29\%$ BD-Rate savings on average under LDB and RA configurations, respectively. Moreover, superior coding performance can be achieved on high resolution videos, as more valid samples are provided for modelling, leading to higher fitting accuracy. Moreover, when compared with the fixed-QP coding scheme, the proposed rate control scheme brings 2.96\% BD-Rate savings under LDB configuration and 4.36\% BD-Rate loss under RA configuration. It is worthy to mention that both of the proposed and the default rate control algorithms are capable of improving the coding performance under LDB configuration.
The proposed rate control scheme could provide more efficient coding parameters, leading to further improvement of coding gains. 
However, the rate control may degrade the RD performance under RA configuration compared with the fixed-QP coding.
Furthermore, we exemplified RD curves of sequence ``RaceHorses'' from class C in Fig.~\ref{fig:RD_compare2} from which the RD performance improvement brought by the proposed algorithm can be observed. The encoding complexity of the proposed rate control scheme is tabulated in the last row of Table~\ref{tab:overall_BDRate}. 
The proposed algorithm moderately increases the computational complexity by around 20\%
compared with the default rate control algorithm and the original anchor.

Table~\ref{tab:bitrate_error} illustrates the average bit-rate error of the proposed rate control and the default rate control under LDB and RA configurations where the proposed scheme achieves lower bit-rate error.
Moreover, the bit-rate errors regarding test sequences ``BasketballDrive'' and ``BQMall'' with respect to different target bit-rates under LDB and RA configurations are shown in Table~\ref{tab:example_Lowdelay} and Table~\ref{tab:example_Randomaccess}. Compared with the default rate control algorithm, the propose rate control achieves substantially smaller bit-rate error under RA configuration with varied target bit-rates. Moreover, for LDB configuration, a similar level of the bit-rate error regarding the default rate control and the proposed rate control can be observed.

To further demonstrate the benefits of the proposed method, the PSNR and the output bit-rate of individual frame in sequence ``BasketballDrill'' are extracted under RA configuration where the target bit-rate is set to 2856 kbps.
We illustrate the instant PSNR and the output bit-rate from POC 60 to POC 92 in Fig.~\ref{fig:RD_compare1} with the cooperation of the default rate control scheme and the proposed scheme.

It can be observed that the proposed rate control scheme reveals a similar trend to the default scheme regarding the output coding bits in varied frames, wherein the key frames such as POC 64 and POC 80 could enjoy more bits. 
Moreover, owing to the proper bit allocation, the proposed scheme achieves superior PSNR performance compared with the default rate control scheme, especially in terms of the key frames, leading to overall performance improvement. Fig.~\ref{fig:sequence} illustrates the output bits by per-second for sequence ``RitualDance'' under LDB and RA configurations, where the associated target bit-rate is set to 2876 kbps and 2467 kbps. Compared with the default rate-control algorithm, the output bit-rates are more stable when employing the proposed rate control schemes.

\section{Conclusion}
In this paper, we propose a novel rate control algorithm for VVC based on an improved Cauchy distribution, which achieves superior compression performance compared with the default frame-level rate control algorithm in VTM-3.0. 
Based on the proposed distribution model, we theoretically derive R-Q and D-Q models which are demonstrated to realize higher modelling accuracy regarding the RD characteristics of diversified video contents.
Furthermore, we explore the frame dependency between different temporal layers, with which an adaptive bit allocation scheme is established for optimal bit allocation. Compared with the VVC rate control algorithm, owing to proper bit allocation and accurate Q-$\lambda$ relationship, the proposed algorithm can achieve 1.03\% BD-Rate savings under LDB configuration and 1.29\% BD-Rate savings under RA configuration. Moreover, with LDB configuration, the proposed algorithm outperforms the fixed-QP coding scheme, where 2.96\% BD-Rate savings can be achieved. These results provide meaningful evidence regarding the effectiveness of the proposed rate control algorithm.

\appendix[Proof of Eqn.~(\ref{4.7})]
According to Eqn.~(\ref{4.6}), we assume, 
\begin{align}
\label{ap_1}
    \sum_{i=1}^{N_f} J_{ex}^i = \sum_{i=1}^{N_f} \sum_j J_{ex}^i(\hat{D_j}(Q_j)-D_p^{j_p}),
\end{align}
where $j$ is the index of reference list regarding the frame $i$. $Q_j$ denotes the quantization step size of the reference frame $j$. We set
\begin{align}
    A^i_j = J_{ex}^i(\hat{D_j}(Q_j)-D_p^{j_p}),
\end{align}
where $A^i_j$ means external RD cost of frame $i$, which is caused from frame $j$'s fluctuation. Then we expand $j$ to the whole GOP. By setting $A^i_j$ equaling to zero, if frame $j$ is not in frame $i$'s reference list, Eqn.~(\ref{ap_1}) can be written as,
\begin{align}
\label{ap_2}
    \sum_{i=1}^{N_f} J_{ex}^i =& \sum_{i=1}^{N_f} \sum_{j=1}^{N_f} A^i_j=\sum \begin{bmatrix}
A_1^1 & \dots & A_{N_f}^1\\
 \vdots & \ddots & \vdots\\
A_1^{N_f} &\dots & A_{N_f}^{N_f}
\end{bmatrix}_{N_f \times N_f}\\\nonumber
=&\sum_{j=1}^{N_f}(A^1_j+A^2_j+\dots+A^{N_f}_j)=\sum_{j=1}^{N_f}\sum_{i=1}^{N_f} A^i_j.
\end{align}
Based on our assumption, if frame $j$ is not in frame $i$'s reference list, $A_j^i$ equals to zero. Eqn.~(\ref{ap_2}) can be written as,
\begin{align}
    \sum_{i=1}^{N_f} J_{ex}^i =& \sum_{j=1}^{N_f}\sum_{i=1}^{N_f} A^i_j = \sum_{j=1}^{N_f}\sum_k A^k_j\\\nonumber
    =&\sum_{j=1}^{N_f} \sum_k J_{ex}^k(\hat{D_j}(Q_j)-D_p^{j_p}).
\end{align}
Herein, $k$ is the list of frames which employ frame $j$ as reference frame.
\ifCLASSOPTIONcaptionsoff
  \newpage
\fi

\bibliographystyle{IEEEtran}

\small
\bibliography{bibtex/bib/egbib.bib}
\end{document}